\documentclass[fleqn,usenatbib]{mnras}

\usepackage{newtxtext,newtxmath}

\usepackage[T1]{fontenc}

\DeclareRobustCommand{\VAN}[3]{#2}
\let\VANthebibliography\thebibliography
\def\thebibliography{\DeclareRobustCommand{\VAN}[3]{##3}\VANthebibliography}

\usepackage{graphicx}	
\usepackage{amsmath}	

\usepackage{amssymb}	
\usepackage{upgreek}
\usepackage{pdflscape}

\title[HI disc size and star formation in Hydra I]{WALLABY Pilot Survey: H\,\textsc{i} gas disc truncation and star formation of galaxies falling into the Hydra I cluster}

\author[T.N.~Reynolds, et al.]{T.N.~Reynolds,$^{1,2}$\thanks{tristan.reynolds@uwa.edu.au} 
B.~Catinella,$^{1,2}$
L.~Cortese,$^{1,2}$
T.~Westmeier,$^{1,2}$
G.R.~Meurer,$^{1}$
L.~Shao,$^{3}$
\newauthor
D.~Obreschkow,$^{1,2}$
J.~Rom{\'a}n,$^{4,5}$
L.~Verdes-Montenegro,$^{6}$
N.~Deg,$^{7}$
H.~D{\'e}nes,$^{8}$
B.-Q.~For,$^{1,2}$
\newauthor
D.~Kleiner,$^{9}$
B.S.~Koribalski,$^{10,11}$
K.~Lee-Waddell,$^{1,12}$
C.~Murugeshan,$^{12,2}$
S.-H.~Oh,$^{13}$
J.~Rhee,$^{1,2}$
\newauthor
K.~Spekkens,$^{14}$
L.~Staveley-Smith,$^{1,2}$
A.R.H.~Stevens,$^{1,2}$
J.M.~van~der~Hulst,$^{15}$
J.~Wang,$^{16}$
\newauthor
O.I.~Wong,$^{12,1,2}$
B.W.~Holwerda,$^{17}$
A.~Bosma,$^{18}$
J.P.~Madrid,$^{19}$
K.~Bekki$^{1}$
\\
$^1$International Centre for Radio Astronomy Research (ICRAR), The University of Western Australia, 35 Stirling Hwy, Crawley, WA, 6009, Australia\\
$^2$ARC Centre of Excellence for All Sky Astrophysics in 3 Dimensions (ASTRO 3D), Australia \\
$^{3}$National Astronomical Observatories, Chinese Academy of Sciences, 20A Datun Road, Chaoyang District, Beijing, China \\
$^{4}$Instituto de Astrof\'{\i}sica de Canarias, c/ V\'{\i}a L\'actea s/n, E-38205, La Laguna, Tenerife, Spain \\
$^{5}$Departamento de Astrof\'{\i}sica, Universidad de La Laguna, E-38206, La Laguna, Tenerife, Spain \\
$^{6}$Instituto de Astrof\'{\i}sica de Andaluc\'{\i}a, CSIC, Glorieta de la Astronom\'{\i}a, E-18080, Granada, Spain \\
$^{7}$Department of Physics, Engineering Physics, and Astronomy, Queen's University, Kingston, ON, K7L 3N6, Canada \\
$^{8}$ASTRON - The Netherlands Institute for Radio Astronomy, 7991 PD Dwingeloo, The Netherlands \\
$^{9}$INAF – Osservatorio Astronomico di Cagliari, Via della Scienza 5, 09047 Selargius, CA, Italy \\
$^{10}$CSIRO Astronomy and Space Science, Australia Telescope National Facility, P.O. Box 76, Epping NSW 1710, Australia \\
$^{11}$Western Sydney University, Locked Bag 1797, Penrith, NSW 2751, Australia \\
$^{12}$CSIRO Astronomy and Space Science, PO Box 1130, Bentley WA 6102, Australia \\
$^{13}$Department of Physics and Astronomy, Sejong University, 209 Neungdong-ro, Gwangjin-gu, Seoul, Republic of Korea \\
$^{14}$Department of Physics and Space Science Royal Military College of Canada P.O. Box 17000, Station Forces Kingston, ON K7K 7B4, Canada \\
$^{15}$Kapteyn Astronomical Institute, University of Groningen, Landleven 12, 9747AD Groningen, The Netherlands \\
$^{16}$Kavli Institute for Astronomy and Astrophysics, Peking University, Beijing 100871, China \\
$^{17}$Department of Physics and Astronomy, University of Louisville, 40292 KY Louisville, USA \\
$^{18}$Aix Marseille Univ, CNRS, CNES, LAM, Marseille, France \\
$^{19}$The University of Texas Rio Grande Valley, One West University Blvd, Brownsville, TX 78520 USA \\
}

\date{Accepted 2021 November 29. Received 2021 November 22; in original form 2021 October 01}

\pubyear{2021}

\begin{document}
\label{firstpage}
\pagerange{\pageref{firstpage}--\pageref{lastpage}}
\maketitle

\begin{abstract}
We present results from our analysis of the Hydra~I cluster observed in neutral atomic hydrogen (H\,\textsc{i}) as part of the Widefield ASKAP L-band Legacy All-sky Blind Survey (WALLABY). These WALLABY observations cover a 60-square-degree field of view with uniform sensitivity and a spatial resolution of 30\,arcsec. We use these wide-field observations to investigate the effect of galaxy environment on H\,\textsc{i} gas removal and star formation quenching by comparing the properties of cluster, infall and field galaxies extending up to $\sim5R_{200}$ from the cluster centre. We find a sharp decrease in the H\,\textsc{i}-detected fraction of infalling galaxies at a projected distance of $\sim1.5R_{200}$ from the cluster centre from $\sim0.85\%$ to $\sim0.35\%$. We see evidence for the environment removing gas from the outskirts of H\,\textsc{i}-detected cluster and infall galaxies through the decrease in the H\,\textsc{i} to $r$-band optical disc diameter ratio. These galaxies lie on the star forming main sequence, indicating that gas removal is not yet affecting the inner star-forming discs and is limited to the galaxy outskirts. Although we do not detect galaxies undergoing galaxy-wide quenching, we do observe a reduction in recent star formation in the outer disc of cluster galaxies, which is likely due to the smaller gas reservoirs present beyond the optical radius in these galaxies. Stacking of H\,\textsc{i} non-detections with H\,\textsc{i} masses below $M_{\rm{HI}}\lesssim10^{8.4}\,\rm{M}_{\odot}$ will be required to probe the H\,\textsc{i} of galaxies undergoing quenching at distances $\gtrsim60$\,Mpc with WALLABY.
\end{abstract}




\section{Introduction}
\label{sec:intro}

The environment in which a galaxy resides has a large effect on its observed properties. This is clearly demonstrated by the morphology-density relation \citep[e.g.][]{Oemler1974, Dressler1980}, in which the fraction of late-type galaxies decreases (and early-type galaxies increase) with increasing galaxy number density (i.e.\ moving from galaxies located in the field to the centre of clusters). The environments with the highest galaxy number and intergalactic medium (IGM) densities are galaxy clusters, containing hundreds to thousands of galaxies. 

The fraction of passive, non-star-forming galaxies is also found to be larger in clusters than in the field \citep[e.g.][]{Kauffmann2004, PintosCastro2019}. Some of the apparent relation between the level of star formation and environment is due to galaxy morphology, as clusters contain more early-type galaxies, which tend to be quenched. However, galaxies in clusters are found to be less star-forming than isolated field galaxies with similar stellar mass and bulge fraction \citep[e.g.][]{Balogh1997, Balogh1998}. In the Virgo cluster, the reduction in the total star formation traced by H$\upalpha$ emission of spiral galaxies is found to be caused by these galaxies having truncated star-forming discs compared to similar field galaxies \citep{Koopmann2004a, Koopmann2004b}. 

Star formation is connected to the neutral atomic hydrogen (H\,\textsc{i}) gas content of galaxies, as H\,\textsc{i} provides a potential reservoir for future star formation through its conversion to molecular gas, H$_2$ \citep[e.g.][]{Bigiel2008,Leroy2008}. For cluster galaxies, the size of the star-forming and optical discs estimated using a number of star formation indicators \citep[e.g. ultraviolet, H$\upalpha$ and 24$\upmu$m emission,][]{Cortese2012, Fossati2013, Finn2018} shows a strong correlation with HI deficiency. Late-type, spiral galaxies are found to be more gas-poor in cluster and group environments than in the field \citep[see the review by][and references therein]{Cortese2021} and the fraction of galaxies detected in H\,\textsc{i} decreases with increasing density \citep[e.g.][]{Hess2013}. Galaxies in clusters are also found to have truncated H\,\textsc{i} discs \citep[e.g.][]{Cayatte1990, BravoAlfaro2000, Yoon2017}. Disc truncation is not only observed in H\,\textsc{i} gas. In the Virgo cluster, truncated discs have been observed also in the molecular gas \citep[e.g.][]{Boselli2014a}, in the dust \citep{Cortese2010}, and in the ionised gas disc \citep[e.g.][]{Koopmann2004b, Boselli2006, Boselli2015}. This results in a reduction in fuel from which to form stars and may lead to the quenching of star formation. \cite{Boselli2016} find a decrease in the star formation of H\,\textsc{i}-deficient Virgo cluster galaxies. \cite{Yoon2017} also find that Virgo cluster galaxies which are H\,\textsc{i}-deficient are also less star-forming and have their star formation segregated towards their centres. 

The lack of H\,\textsc{i} in cluster galaxies can be caused by a combination of the cessation of gas inflows replenishing a galaxy's H\,\textsc{i} reservoir once the current gas content is depleted and/or the removal of gas through stripping. There are both gravitational and hydrodynamical environmental processes that can lead to H\,\textsc{i} deficient cluster galaxies. Galaxy-galaxy interactions include tidal stripping \citep[e.g.][]{Moore1999}, harassment \citep[e.g.][]{Moore1996, Moore1998} and mergers \citep[e.g.][]{Toomre1972}. Mechanisms acting between the IGM and a galaxy include ram pressure stripping \citep[e.g.][]{Gunn1972}, viscous stripping \citep[e.g.][]{Nulsen1982, Quilis2000}, starvation \citep[e.g.][]{Larson1980} and thermal evaporation \citep[e.g.][]{Cowie1977}. 

A galaxy's H\,\textsc{i} gas normally extends to the outskirts of a galaxy and is generally the first easily detectable component of a galaxy to be influenced by the environment. Thus the emission from H\,\textsc{i} serves as a sensitive probe of the impact of the environment in which the galaxy resides. While blind, single dish surveys such as HIPASS and ALFALFA \citep[][respectively]{Meyer2004, Haynes2018} have observed clusters in H\,\textsc{i}, they lack spatial resolution and are limited to measuring integrated H\,\textsc{i} properties. A number of nearby clusters have been the subject of targeted \citep[e.g.\ Coma, Ursa Major and Virgo,][respectively]{BravoAlfaro2000, Verheijen2001, Chung2009} and blind \citep[e.g.\ Abell\,2626 and Fornax,][respectively]{Healy2021b, Loni2021} interferometric surveys and have provided spatially resolved H\,\textsc{i} information. However, these surveys have been limited to modest fields of view (e.g.\ $\sim1$--$2$ virial radii) due to the small field of view and long integration times required for traditional interferometric observations.

The Widefield ASKAP L-band Legacy All-sky Blind Survey \citep[WALLABY,][]{Koribalski2020} on the Australian Square Kilometre Array Pathfinder \citep[ASKAP,][]{Johnston2008, Hotan2021} is beginning to change this. WALLABY aims to cover 3/4 of the sky up to $\delta=+30^{\circ}$ and detect H\,\textsc{i} emission in $\sim500\,000$ galaxies of which $\sim5\,000$ will be spatially resolved. ASKAP is fitted with phase array feed \citep[PAF,][]{DeBoer2009, Hampson2012, Hotan2014, Schinckel2016} receivers, which provide a 30-square-degree instantaneous field of view footprint on the sky. Full sensitivity across the field of view is reached by interleaving two ASKAP footprints diagonally offset by $\sim0.64^{\circ}$ (creating an ASKAP tile) and WALLABY is able to reach its nominal sensitivity of 1.6\,mJy per beam per 4\,km\,s$^{-1}$ channel with an integration time of 16\,h (8\,h per footprint). 

Prior to the full survey commencing, WALLABY is currently undertaking a pilot survey of a number of 30-square-degree tiles. This includes a 60-square-degree field comprising two adjacent ASKAP tiles covering the Hydra~I cluster and extending out to $\sim5R_{200}$ to the west of the cluster. These data have been used for the detailed study of individual objects \citep[e.g.\ ESO\,501$-$G075,][]{Reynolds2021} and larger population studies (e.g.\ ram pressure of galaxies within $\sim2.5R_{200}$, \citeauthor{Wang2021} \citeyear{Wang2021}). 

\subsection{The Hydra I Cluster}
\label{s-sec:hydra_cluster}

The Hydra~I cluster \citep[`Abell\,1060' in][]{Abell1958} is centred on $\alpha,\delta=10$:36:41.8, $-27$:31:28 (J2000) and has a heliocentric recessional velocity of c$z\sim3\,780$\,km\,s$^{-1}$ \citep{Struble1999, Panagoulia2014}. The centre of the Hydra~I cluster is dominated by NGC\,3311 and NGC\,3309, two giant elliptical galaxies (indicated by the grey star in Figure~\ref{fig:sky_pos}). The systemic velocity in the CMB reference frame is c$z=4\,120$\,km\,s$^{-1}$. This gives a luminosity distance of $D_{\mathrm{L}}=61$\,Mpc, which we adopt as the distance of Hydra~I throughout this work, and is in good agreement with the redshift-independent distance of 59\,Mpc \citep{Jorgensen1996}. Hydra~I has a velocity dispersion of $\sigma_{\rm{disp}}=676\pm35$\,km\,s$^{-1}$ \citep{Richter1982}. We adopt the cluster virial radius of $R_{200}\sim1.44\pm0.08$\,Mpc from \cite{Reiprich2002} ($\sim1.35^{\circ}$ projected on the sky at 61\,Mpc). The corresponding cluster mass within $R_{200}$ is $M_{200}=(3.13\pm0.50)\times10^{14}\,\rm{M}_{\odot}$ \citep{Reiprich2002}. \cite{Solanes2001} find Hydra~I to be H\,\textsc{i}-deficient based on a sample of 96 galaxies within $5R_{\rm{Abell}}$ (Abell radius, $R_{\rm{Abell}}=2.16^{\circ}$) of the cluster centre, 20 of which are within $R_{\rm{Abell}}$. 

There is tension in the literature over the dynamical state of Hydra~I. \cite{Fitchett1988} find that the velocity distribution of galaxies within $\sim40$\,arcmin of the centre of Hydra~I is non-Gaussian and propose clumpy substructure as the cause for this non-Gaussian distribution. Hydra~I is found to be slightly disturbed with a substructure in the process of falling into the cluster in the foreground \citep{LimaDias2021}. \cite{Arnaboldi2012} and \cite{Barbosa2018} find evidence for a recent sub-cluster merger event. These results suggest that Hydra~I is not yet virialised. However, other studies have found that Hydra~I has a fairly homogeneous X-ray distribution, which suggests that the cluster is relaxed and has not undergone any recent mergers \citep[e.g.][]{Fitchett1988, Hayakawa2004, Lokas2006}, although a small offset between the X-ray centre of the cluster core and the centre of the brightest galaxy NGC\,3311 has also been found \citep{Hayakawa2006}. 

Hydra~I is similar to the Virgo cluster, which has a virial mass of  $M_{\rm{vir}}=(6.3\pm0.9)\times10^{14}\,\rm{M}_{\odot}$ and velocity dispersion of $\sigma_{\rm{disp}}=638\pm35$\,km\,s$^{-1}$ \citep{Kashibadze2020}, but is $\sim3.5$ times more distant (61 vs 16.5\,Mpc, \citeauthor{Mei2007} \citeyear{Mei2007}). At this increased distance, a single ASKAP tile centred on Hydra~I covers the cluster out to $\sim2.5R_{200}$ (at the Virgo cluster distance this would cover out to $\sim R_{200}$). This provides a more complete picture of the cluster environment by including the infall region surrounding the cluster core and enables us to study the effect of pre-processing on gas removal and star formation quenching in and around the cluster. 

In this work, we investigate gas removal and star formation in the Hydra~I cluster using wide-field, high spatial resolution WALLABY observations that cover 60 square degrees, going out to $\sim5R_{200}$ from the cluster centre, by comparing the H\,\textsc{i} to optical disc diameter ratio and star formation rate of cluster and infall galaxies with a control sample of field galaxies. We present the data we use for this work and derive physical quantities in Section~\ref{sec:data}. We present and discuss our results in Sections~\ref{sec:results} and \ref{sec:discussion} and summarise our conclusions in Section~\ref{sec:conclusion}. Throughout, we adopt optical velocities (c$z$) in the heliocentric reference frame, the AB magnitude convention and we assume a flat $\Lambda$CDM cosmology with $H_0=67.7$\,km\,s$^{-1}$\,Mpc$^{-1}$ \citep{Planck2016}.

\section{Data}
\label{sec:data}

\subsection{WALLABY Observations and Sample Selection}
\label{s-sec:hi_data}

The WALLABY pilot survey observations of the Hydra field cover 60 square degrees extending over 10:03:$00<\alpha\,\rm{(J2000)}<10$:53:01 and $-30$:30:$00<\delta\,\rm{(J2000)}<-24$:30:00. We refer the reader to the following WALLABY publications for details of the WALLABY pilot survey observations of the Hydra~I cluster: \citet{Reynolds2021,Wang2021} and ASKAP/WALLABY data reduction process: \citet{Elagali2019, For2019, Kleiner2019, Lee-Waddell2019, Reynolds2019}. The final H\,\textsc{i} spectral line cube has a synthesised beam size of $30\times30$\,arcsec, spectral resolution of 4\,km\,s$^{-1}$ and an average root-mean-square (rms) sensitivity of $\sim 2.0$\,mJy per beam per 4\,km\,s$^{-1}$ over the cube. We note that the average sensitivity across the cube is higher than the nominal sensitivity due to a bandpass ripple artefact in the spectral line cube. The bandpass ripple is due to the ASKAP On-Dish Calibrators (ODC).\footnote{ASKAP Update, November 2020 \url{https://www.atnf.csiro.au/projects/askap/ASKAP_com_update_v44.pdf}} Three beams (located in the corners of individual footprints), which are strongly affected by the ODC ripples, were removed from the 144 beams making up the 60-square-degree mosaic prior to running the H\,\textsc{i} source finding. The ASKAP operations team have since turned off the ODCs and this will not impact future spectral line observations.

We use the Source Finding Application 2 \citep[SoFiA 2,][]{Serra2015a,Westmeier2021} to detect sources of H\,\textsc{i} emission across a redshift range of c$z\sim500$--25\,000\,km\,s$^{-1}$ using the SoFiA smooth$+$clip (S$+$C) finder. We first apply preconditioning to the data cube using the following steps: (1) multiplication of the data by the square root of the weights cube produced by the ASKAP pipeline; (2) local noise normalisation across a running window of $51\times51$ spatial pixels and 51 spectral channels in size; (3) auto-flagging of bad data using an internal threshold of 5. For the S$+$C finder, we set Gaussian spatial filter sizes of 0, 5 and 10 pixels and spectral boxcar filter sizes of 0, 3, 7 and 15 channels. We set the source-finding threshold to $3.5\sigma$ with a replacement value of $2\sigma$. Detections are linked across a spatial and spectral radius of 2 pixels/channels with a minimum size requirement for reliable source of 8 spatial pixels and 5 spectral channels. SoFiA’s reliability filter is then applied to remove all detections with a reliability below 0.8, using a Gaussian kernel density estimator of size 0.4 times the covariance. All remaining sources are then parameterised, assuming a restoring beam size of $\sim30$\,arcsec for all integrated flux measurements. 

After removing artefacts (i.e.\ related to the bandpass ripple) from the detection catalogue, the final catalogue contains 272 H\,\textsc{i} detections with integrated signal-to-noise ratios $\rm{SNR}\gtrsim5$ (i.e.\ $\rm{SNR}=S_{\rm{sum}}/\sigma_{S_{\rm{sum}}}$, where $S_{\rm{sum}}$ is the integrated flux from the SoFiA source mask and $\sigma_{S_{\rm{sum}}}$ is the statistical uncertainty in the integrated flux). As Hydra~I has a systemic velocity of c$z\sim3\,780$\,km\,s$^{-1}$, we select a subsample of the H\,\textsc{i} detections below c$z<7\,000$\,km\,s$^{-1}$ (i.e.\ detections for which the H\,\textsc{i} mass sensitivity will be similar to cluster members). This is to avoid the bias towards high stellar mass and gas-rich galaxies that WALLABY will detect at higher redshifts. We also exclude five detections of interacting systems contained within a H\,\textsc{i} envelope whose projected angular separation is less than the ASKAP beam size, as we are unable to measure individual H\,\textsc{i} properties for each galaxy in these systems. These cuts result in a sample of 145 individual galaxies with detected H\,\textsc{i} emission and systemic velocities c$z<7\,000$\,km\,s$^{-1}$. 

\begin{figure*}
	\includegraphics[width=\textwidth]{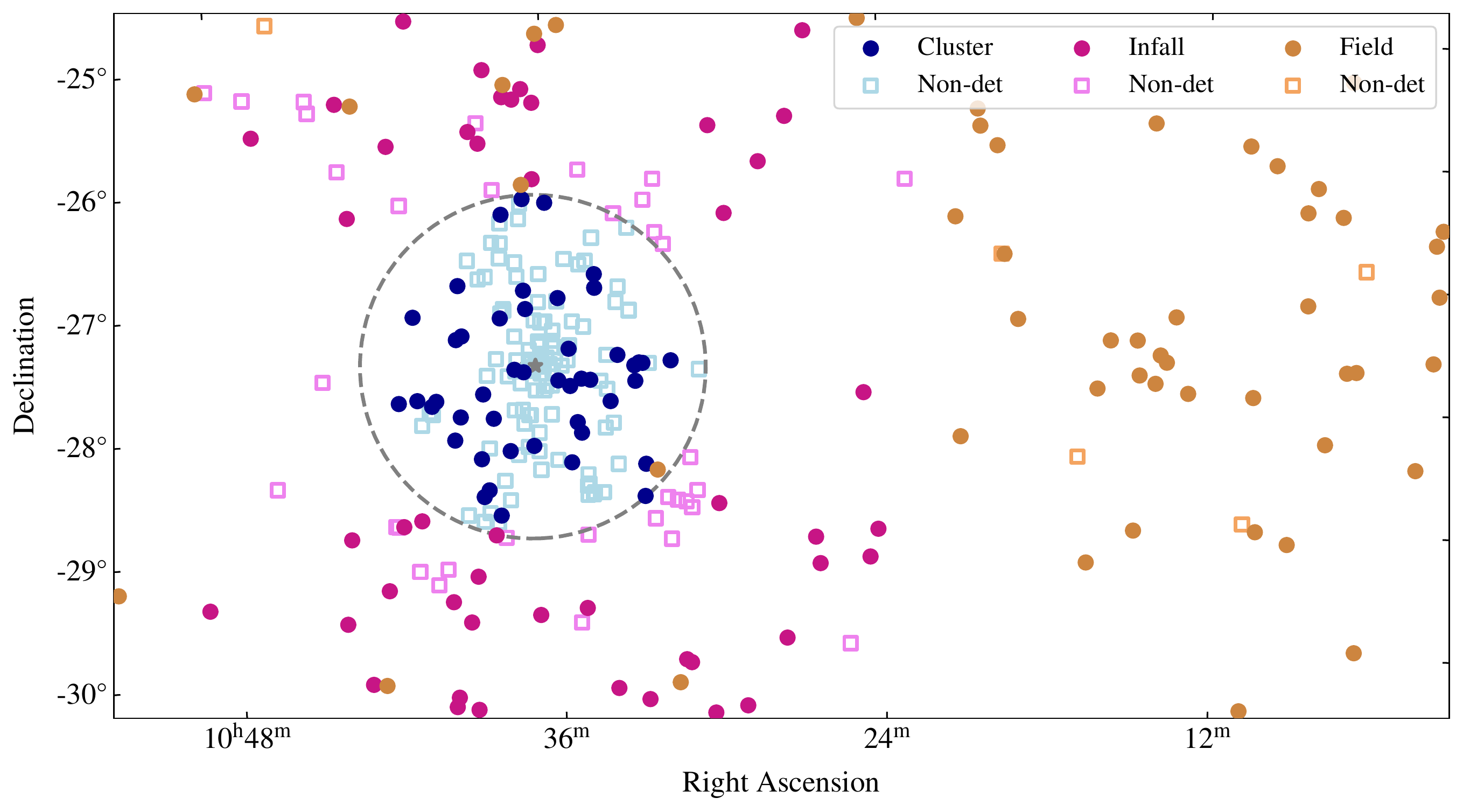}
    \caption{Position of galaxies with c$z<7\,000$\,km\,s$^{-1}$ in the 60-square-degree WALLABY footprint. H\,\textsc{i} detections and non-detections are indicated by the filled circles and unfilled squares, respectively. The cluster, infall and field galaxy populations defined using the phase space diagram in Figure~\ref{fig:hydra_psd} are coloured blue, purple and orange, respectively. The grey dashed circle indicates $R_{200}$ (i.e.\ the virial radius) of the Hydra~I cluster ($1.44$\,Mpc) and the grey star indicate the positions of NGC\,3311 and NGC\,3309 (i.e.\ the approximate centre of the cluster).}
    \label{fig:sky_pos}
\end{figure*}

As a blind H\,\textsc{i} survey, WALLABY is most sensitive to gas-rich galaxies and will detect few gas-poor galaxies (predominantly early-types), which are the dominant type of galaxy found in clusters. Thus, H\,\textsc{i} detections alone do not provide a complete galaxy sample for probing the effect of the environment on H\,\textsc{i} content and star formation \citep[e.g.\ demonstrated by][using ALFALFA\footnote{Arecibo Legacy Fast ALFA Survey \citep{Giovanelli2005}} data]{Yoon2015}. We use the 6dF Galaxy Survey \citep[6dFGS,][]{Jones2009} to identify galaxies within the Hydra field footprint and with systemic velocities c$z<7\,000$\,km\,s$^{-1}$ without detected H\,\textsc{i} emission. The 6dFGS survey is complete to near infrared magnitudes of 12.65, 12.95 and 13.75 in the $K$-, $H$- and $J$-bands and at the distance of Hydra~I detects galaxies with stellar masses $\gtrsim10^9\,\rm{M}_{\odot}$. Hence, due to the sensitivity limits of WALLABY and the magnitude limits of 6dFGS, our sample is biased against low mass dwarf galaxies (e.g.\ low-surface brightness and ultra-diffuse galaxies detected in Hydra~I by \citeauthor{Iodice2020} \citeyear{Iodice2020}). We derive H\,\textsc{i} mass upper limits for these non-detections in Section~\ref{ss-sec:hi_mass}. Figure~\ref{fig:sky_pos} shows the projected sky distribution of H\,\textsc{i}-detected galaxies (filled circles) and galaxies from 6dFGS not detected in H\,\textsc{i} (unfilled squares) with c$z<7\,000$\,km\,s$^{-1}$. The Hydra~I cluster lies in the eastern half of the observed footprint (the dashed circle indicates $R_{200}$) and shows a clear concentration of galaxies with H\,\textsc{i} non-detections dominating within $R_{200}$.

\begin{figure}
	\includegraphics[width=\columnwidth]{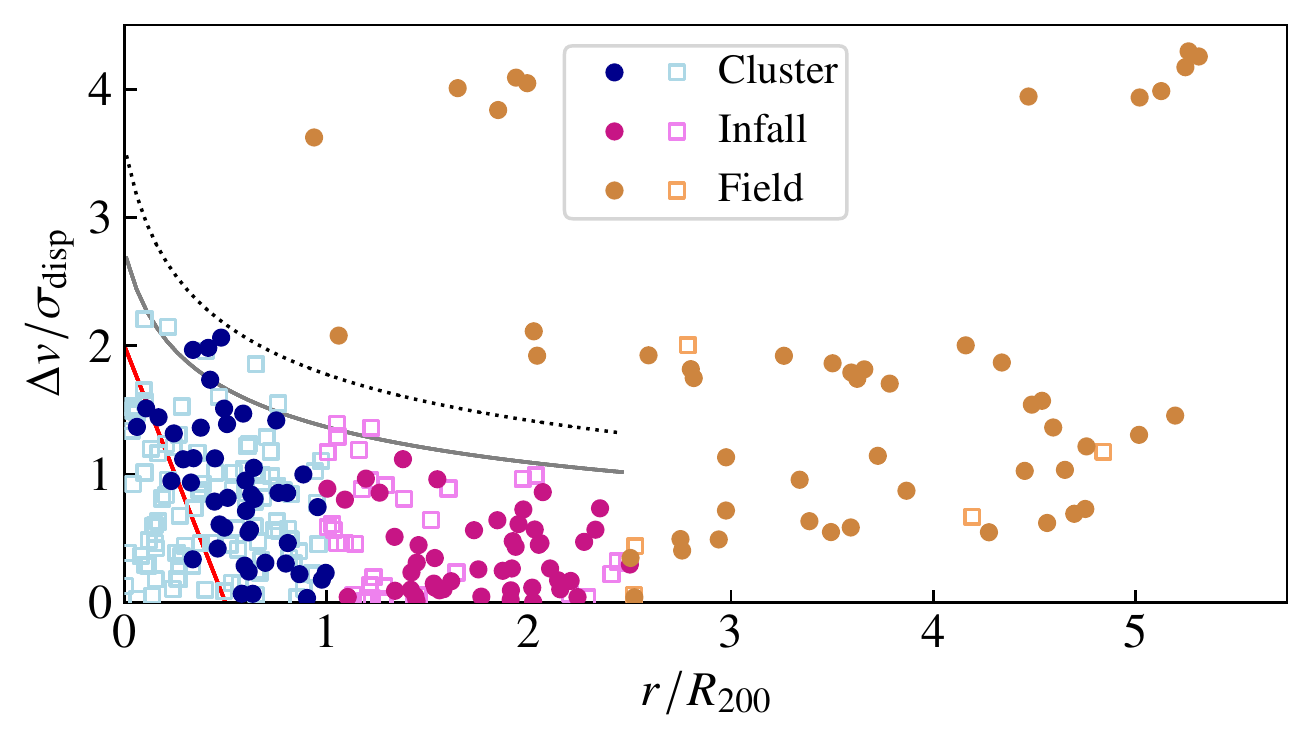}
    \caption{Hydra~I cluster phase space diagram used to classify galaxies into the three populations: cluster, infall and field (blue, purple and orange, respectively). Filled circles and unfilled squares indicate H\,\textsc{i} detections and non-detections respectively. The virialised region defined by \citet{Rhee2017} lies to the left of the red diagonal line. The escape velocity curve for Hydra~I is indicated by the solid grey curve with $3\sigma_{\rm{esc}}$ above the escape velocity curve indicated by the dotted grey curve.}
    \label{fig:hydra_psd}
\end{figure}

We classify galaxies within the Hydra field footprint as cluster, infall or field galaxies based on their location on a phase space diagram \citep[e.g.][]{Rhee2017} of the Hydra~I cluster (Figure~\ref{fig:hydra_psd}). Cluster galaxies have projected distances from the cluster centre of $r<R_{200}$ and velocities relative to the cluster systemic velocity, $\Delta v/\sigma_{\rm{disp}}<3\sigma_{\rm{esc}}$, where $\sigma_{\rm{esc}}$ is the uncertainty in the escape velocity, above the cluster escape velocity curve (i.e.\ below the dotted black curve). We calculate the escape velocity and $\sigma_{\rm{esc}}$ following equations~1--4 from \cite{Rhee2017} using $R_{200}$, $M_{200}$ and $\sigma_{\rm{disp}}$ given in Section~\ref{s-sec:hydra_cluster}. Infall galaxies are defined by $R_{200}<r<2.5R_{200}$ and $\Delta v/\sigma_{\rm{disp}}<3\sigma_{\rm{esc}}$ above the cluster escape velocity curve. All other galaxies are classified as field galaxies. We detect significantly more galaxies in H\,\textsc{i} with WALLABY compared to those listed by \cite{Solanes2001}: 44 within $1.35^{\circ}$ vs 20 within $2.16^{\circ}$ and 129 within $\sim6.75^{\circ}$ vs 96 within $10.8^{\circ}$ (also noting that the WALLABY observations extend out $\sim6.75^{\circ}$ to the West of the cluster and $\sim2.5^{\circ}$ in all other directions). We also detect $\sim6$ times more galaxies than are included in the HIPASS catalogue within this region of sky (23 individual galaxies with integrated signal-to-noise ratio $\rm{SNR}>5$, \citeauthor{Barnes2001} \citeyear{Barnes2001}; \citeauthor{Meyer2004} \citeyear{Meyer2004}) due to the increased sensitivity of WALLABY ($\sim2$ vs $\sim13$\,mJy\,beam$^{-1}$). We tabulate the environment classifications and the galaxy properties we estimate/compute in Section~\ref{sec:data} (i.e.\ H\,\textsc{i} and stellar mass, H\,\textsc{i}, $r$-band and NUV disc diameters and star formation rates) in Appendix~\ref{appendix:data_tables}, with the full tables available as supplementary online material.

\subsubsection{H\textsc{i} Mass}
\label{ss-sec:hi_mass}

In this work, we use two H\,\textsc{i} properties: the total H\,\textsc{i} mass and the size of the H\,\textsc{i} disc. We use the integrated fluxes and integrated intensity (moment 0) maps to measure these quantities for the H\,\textsc{i} detections and derive upper limits for the H\,\textsc{i} non-detections.

For galaxies detected in H\,\textsc{i}, we calculate the H\,\textsc{i} mass, $M_{\rm{HI}}$, using equation~48 from \cite{Meyer2017},
\begin{equation}
    \frac{M_{\rm{HI}}}{\rm{M}_{\odot}} = 49.7 \left(\frac{D_{\rm{L}}}{\rm{Mpc}}\right)^{2} \left(\frac{S}{\rm{Jy\,Hz}}\right),
    \label{equ:mhi}
\end{equation}
where $D_{\rm{L}}$ is the luminosity distance and $S$ is the integrated flux. We can place an upper limit on $M_{\rm{HI}}$ for those galaxies not detected with WALLABY using the minimum SNR of our H\,\textsc{i} detections ($\rm{SNR}=5$) and flux density sensitivity. We derive the theoretical H\,\textsc{i} mass sensitivity of WALLABY at the distance of the Hydra~I cluster (61\,Mpc) using equation~157 for the SNR for a given set of observational parameters from \cite{Meyer2017},
\begin{multline}
    \frac{M_{\rm{HI}}}{\rm{M}_{\odot}} = \left(\frac{\rm{SNR}}{2.92 \times 10^{-4}}\right) (1+z)^{-1/2} \left(\frac{D_{\rm{L}}}{\rm{Mpc}}\right)^{2}  \left(\frac{\Delta V}{\rm{km\,s}^{-1}}\right)^{1/2} \\
    \times  \left(\frac{\Delta \nu_{\mathrm{chan}}}{\rm{Hz}}\right)^{1/2} \left(\frac{\sigma}{\rm{Jy}}\right) \left(1+\frac{A_{\rm{galaxy}}}{A_{\rm{beam}}}\right)^{1/2},
    \label{equ:mhi_limit}
\end{multline}
where $\Delta V$ is the spectral line width, $\Delta \nu_{\mathrm{chan}}$ is the channel width, $\sigma$ is the channel noise and $A_{\rm{galaxy}}$ and $A_{\rm{beam}}$ are the areas of the galaxy and synthesised beam. This equation reduces to the form for a point source when $A_{\rm{galaxy}}\ll A_{\rm{beam}}$ and the $(1+A_{\rm{galaxy}}/A_{\rm{beam}})^{1/2}$ term reduces to unity. We assume a fixed distance of $D_{\rm{L}}=61$\,Mpc for cluster and infall galaxies and calculate $D_{\rm{L}}$ from the individual redshifts for our field population. We set the velocity width to $\Delta V=200$, 300\,km\,s$^{-1}$ for galaxies with $M_*<10^{10}\,\rm{M}_{\odot}$ and $>10^{10}\,\rm{M}_{\odot}$, respectively, (i.e.\ the peaks in the $\Delta V$ distributions for xGASS galaxies in these two mass regimes: GASS-low and GASS, \citeauthor{Catinella2018} \citeyear{Catinella2018}). We assume values of $\rm{SNR}=5$, $\Delta \nu_{\mathrm{chan}}=18\,518$\,Hz, $\sigma=0.002$\,Jy, $A_{\rm{beam}}=30^2\uppi/[4\ln(2)]$ and $A_{\rm{galaxy}}=\uppi \epsilon r^2$, where $r$ is the length of the semi-major axis (radius) in arcseconds and $\epsilon$ is the axis ratio. We estimate $r$ from a linear best fit to the H\,\textsc{i} detections radius vs stellar mass over the range $M_*=10^7$--$10^{10}\,\rm{M}_{\odot}$,
\begin{equation}
    \log\left(\frac{r}{\rm{arcsec}}\right)=0.105 \log\left(\frac{M_*}{\rm{M}_{\odot}}\right) + 0.662.
    \label{equ:hi_radius_mstar}
\end{equation}
The axis ratio is taken to be the average PanSTARRS $r$-band axis ratio from all the H\,\textsc{i} detections and non-detections: $\epsilon=0.59$. At 61\,Mpc, for $\mathrm{SNR}=5$ and stellar mass of $M_{*}=10^9\,\rm{M}_{\odot}$, the H\,\textsc{i} mass sensitivity limit of WALLABY is $M_{\rm{HI}}\sim10^{8.4}\,\rm{M}_{\odot}$ for an unresolved source and $M_{\rm{HI}}\sim10^{8.7}\,\rm{M}_{\odot}$ for a resolved source with a H\,\textsc{i} radius given by Equation~\ref{equ:hi_radius_mstar}. The noise is higher (correspondingly lower H\,\textsc{i} mass sensitivity) for spatially resolved sources as the emission is spread over multiple beams \citep{Duffy2012}.

\subsubsection{H\textsc{i} Diameter}
\label{ss-sec:hi_diameter}

For H\,\textsc{i}-detected galaxies, we measure the diameter of the H\,\textsc{i} disc, $d_{\rm{HI}}$, from the moment 0 (integrated intensity) map at a H\,\textsc{i} surface density of $1\,\rm{M}_{\odot}\,\rm{pc}^{-2}$. Defining a galaxy's H\,\textsc{i} disc diameter at this surface density, \cite{Broeils1997} found a tight ($<0.1$\,dex) correlation with the total H\,\textsc{i} mass. We note there is the potential of a bias of underestimating measured H\,\textsc{i} sizes from interferometric observations due to interferometers potentially missing extended emission due to the short spacing problem \citep[e.g.][]{Braun1985}. To address this, we compare the integrated fluxes and integrated spectra of the 23 galaxies detected with WALLABY that are catalogued in HIPASS \citep{Meyer2004}. We find good agreement between both the integrated fluxes and spectra from WALLABY and HIPASS with a median WALLABY to HIPASS flux ratio of $S_{\rm{int,W}}/S_{\rm{int,H}}\sim1.03$. This indicates that the WALLABY observations are recovering most the H\,\textsc{i} emission and our measured H\,\textsc{i} diameters should be reliable.

We extract the H\,\textsc{i} surface density profile from the moment 0 map by fitting a series of annuli with the centre, position angle and inclination angle defined by a 2-dimensional Gaussian fit to the moment 0 map using the \textsc{miriad} task \textsc{ellint}. We then define the radius at which the surface density drops to $1\,\rm{M}_{\odot}\,\rm{pc}^{-2}$ as the H\,\textsc{i} radius, after converting the profile from Jy\,Hz\,beam$^{-1}$ to $\rm{M}_{\odot}\,\rm{pc}^{-2}$ using equation~82 from \cite{Meyer2017} and correcting the surface density profile for inclination using $\cos(i)$, where $i$ is the inclination angle determined from the 2-dimensional Gaussian fit and we assume optically thin emission. We note that for galaxies with high inclinations and/or that are marginally resolved, the WALLABY synthesised beam will dominate the minor axis size. We correct for this effect by deconvolving the fitted Gaussian by the WALLABY synthesised beam before determining the inclination angle. Assuming the galaxy's H\,\textsc{i} distribution at this surface density is symmetric, the H\,\textsc{i} disc diameter is $d_{\rm{HI},0}=2r_{\rm{HI}}$, where $d_{\rm{HI},0}$ has not been corrected for the effect of the synthesised beam size. Assuming the beam and H\,\textsc{i} disc can both be approximated as Gaussians, this correction takes the form,
\begin{equation}
	d_{\rm{HI}} = \sqrt{d_{\rm{HI},0}^2 - ab},
	\label{equ:dhi_correction}
\end{equation}
where $a$ and $b$ are the synthesised beam major and minor axes ($a=b=30$\,arcsec for WALLABY). H\,\textsc{i} diameters can only be measured for galaxies that are spatially resolved (i.e.\ larger than the synthesised beam). Galaxies with $d_{\rm{HI}}<1.5a$ ($<45$\,arcsec) we classify as unresolved with WALLABY and we consider their $d_{\rm{HI}}$ measurements as upper limits. Of the 129 galaxies detected in H\,\textsc{i}, 10 have $d_{\rm{HI}}<45$\,arcsec and have H\,\textsc{i} masses $M_{\rm{HI}}\lesssim10^{9.04}\,\rm{M}_{\odot}$.

For galaxies not detected in H\,\textsc{i}, we estimate the maximum H\,\textsc{i} disc these galaxies may have based on the H\,\textsc{i} size-mass relation. We use the H\,\textsc{i} size-mass relation from \cite{Wang2016}, $\log(D_{\rm{HI}}/\rm{kpc})=0.506\log(M_{\rm{HI}}/\rm{M}_{\odot}) - 3.293$, and the H\,\textsc{i} mass limit defined previously (Section~\ref{ss-sec:hi_mass}) to estimate upper-limit H\,\textsc{i} disc diameters for the galaxies not detected in H\,\textsc{i}. In dense environments where stripping is occurring, the H\,\textsc{i} size-mass relation holds tight \citep{Wang2016}.

\subsection{PanSTARRS}
\label{s-sec:panstarrs}

We derive stellar masses and measure optical sizes using photometric images from the PanSTARRS \citep{Chambers2016,Flewelling2016} $g$- and $r$-bands. We obtain image cutouts at the position of each H\,\textsc{i} detection and non-detection using the PanSTARRS cutout server.\footnote{\url{https://ps1images.stsci.edu/cgi-bin/ps1cutouts}.} There are 14 galaxies for which the PanSTARRS image cutouts have incomplete coverage and/or artefacts (e.g.\ close to a foreground star), making these galaxies unusable for deriving optical sizes or photometry. We exclude these from our sample (see Table~\ref{table:sample_size} for our final galaxy totals for which we can measure stellar masses and optical diameters).

\begin{table}
	\centering
	\caption{The number of galaxies in each of our three environment classifications (cluster, infall and field) detected in H\,\textsc{i} with WALLABY and H\,\textsc{i} non-detections in 6dFGS.}
	\label{table:sample_size}
	\begin{tabular}{lcccr}
	    \hline
		& Total     & Cluster   & Infall    & Field \\
		\hline
		H\,\textsc{i} Detections       & 129    & 44        & 42        & 43   \\
		H\,\textsc{i} Non-detections   & 142    & 102       & 35        & 5    \\
		\hline
	\end{tabular}
\end{table}

We measure $g$- and $r$-band photometry using the \textsc{python} package \textsc{photutils} in a standard way based on the $r$-band image. We note that PanSTARRS images are already sky background-subracted as part of the PanSTARRS data processing \citep{Magnier2020, Waters2020}. We create a segmentation map using the task \textsc{segmentation} and mask all segments in the $r$-band image other than the target galaxy. We then fit isophotes to the masked image using the task \textsc{isophote}. For PanSTARRS, the image pixel units (ADU) are converted to apparent magnitudes as
\begin{equation}
	m/\mathrm{mag} = 25 + 2.5\log(t) - 2.5\log(y),
	\label{equ:adu_to_mag_panstarrs}
\end{equation}
where $t$ is the total exposure time provided in the image header in seconds and $y$ is the total ADU. The output isophotes provide mean ADU\,pixel$^{-2}$, which we convert to a radial surface brightness in mag\,arcsec$^{-2}$ using Equation~\ref{equ:adu_to_mag_panstarrs} and the PanSTARRS pixel size ($0.2498\times0.2498$\,arcsec\,pixel$^{-2}$). Each isophote is fit to the galaxy independently, without fixing the centre position, position angle or inclination. This allows for the galaxy centre to be automatically determined by the isophote fitting and accurate tracing of changing position angle due to features in the inner disc (e.g.\ bulge and bar). However, a few individual isophotes fail and fit to a different region in the image. We exclude isophotes offset from the galaxy segment centre by $>10$\,arcsec.

We set isophotes with a mean of $<5\sigma_y$ (i.e.\ $<25$\,pixel$^{-2}$, where $\sigma_y\sim5$\,ADU\,pixel$^{-2}$) to the surface brightness of $25.5$\,mag\,arcsec$^{-2}$ (i.e.\ 25\,ADU\,pixel$^{-2}$). The surface brightness limits of the PanSTARRS data in the Hydra~I cluster region are 27.6 and 27.0\,mag\,arcsec$^{-2}$ in the $g$- and $r$-bands, respectively, calculated as $3\sigma$ in $10\times10$\,arcsec boxes following the surface brightness limit definition of \cite{Roman2020}. This surface brightness limit is well above the isophotal contour of 25.5\,mag\,arcsec$^{-2}$ that we use in this work. We measure the total $r$-band magnitude within an aperture defined by the isophote at which the surface brightness drops to 25.0\,mag\,arcsec$^{-2}$ using the \textsc{photutils} task \textsc{aperture\_photometry}. We also use this aperture to extract the total $g$-band image ADU and calculate the total $g$-band magnitude.

\subsubsection{Stellar Mass and $r$-band Diameter}
\label{ss-sec:stellar_mass}

We use the empirical relation from \cite{Taylor2011} and total PanSTARRS $g$- and $r$-band magnitudes to calculate stellar masses,
\begin{multline}
	\log(M_*/\mathrm{M}_{\odot}) = -0.840 + 1.654(g-r) + 0.4(D_{\mathrm{mod}} + M_{\mathrm{sol}} - m) \\ - \log(1+z) - 2\log(h/0.7),
	\label{equ:mstar}
\end{multline}
where $-0.840$ and 1.654 are empirically determined constants based on the $r$-band magnitude and $g-r$ colour from \cite{Zibetti2009}, the $g-r$ colour is in the SDSS photometric system, $m$ is the $r$-band apparent magnitude in the SDSS photometric system, $D_{\mathrm{mod}}$ is the distance modulus (used to convert from apparent to absolute magnitude) and $M_{\mathrm{sol}}=4.64$ is the absolute magnitude of the Sun in the $r$-band \citep{Willmer2018}. We convert magnitudes and colours from PanSTARRS to the SDSS photometric systems using equation~6 from \cite{Tonry2012}. We correct for Galactic extinction in the $g$- and $r$-bands. Assuming the dust extinction law of \cite{Cardelli1989}, the Galactic dust attenuation for a given galaxy is approximated by $A_{\rm{v}} = R_V E(B-V)$, where $R_V=3.793$ and 2.751 for the $g$- and $r$-bands, respectively \citep{Wyder2007}. We obtain the reddening for a given galaxy, $E(B-V)$, from cross-matching our sources with the \textit{GALEX} DR6+7 catalogue \citep{Bianchi2017}. Magnitudes corrected for Galactic dust absorption are then $m_{\rm{cor}}=m_{\rm{obs}} - A_{\rm{v}}$.

Our calculated stellar masses have uncertainties of $\sim0.16$\,dex. We define a galaxy's optical diameter\footnote{We note that optical diameters are often defined at the $B$-band 25\,mag\,arcsec$^{-2}$ surface brightness. However, we define the radius at a surface brightness of 23.5\,mag\,arcsec$^{-2}$ to ensure that we are well above the noise level of the images. Comparing with the 25\,mag\,arcsec$^{-2}$ $B$-band diameters given in the 1989 ESO-Uppsala Catalogue \citep{Lauberts1989, Lauberts2006}, our measured optical diameters are on average 20--30\% smaller.} as the size of the galaxy at which the $r$-band surface brightness reaches 23.5\,mag\,arcsec$^{-2}$. We measure this by interpolating between the two isophotes bridging this surface brightness. Similar to the H\,\textsc{i} diameter, we correct the $r$-band diameter for the size of the point spread function (PSF) using Equation~\ref{equ:dhi_correction} and the PanSTARRS PSF full width at half maximum (FWHM) of $\sim1.25$\,arcsec.

\subsection{GALEX and WISE}
\label{s-sec:wise}

A galaxy's recent ($<100$\,Myr) unobscured star formation is traced by its ultraviolet (UV) luminosity emitted from young stars \citep[e.g.][and references therein]{Kennicutt1998,Kennicutt2012}. The UV emission suffers significant dust attenuation, whereby the UV emission is absorbed by dust and emitted in the infrared \citep[IR, e.g.][]{Buat2005,Hao2011}. As a result, a correction for dust attenuation is required when the UV emission is used to trace a galaxy's star formation rate (SFR). We derive star formation rates using UV and IR photometric images from \textit{GALEX} \citep{Martin2005,Morrissey2007} and WISE \citep{Wright2010}, respectively. We also measure the size of the disc of recent star formation from the \textit{GALEX} UV images.

\subsubsection{GALEX}
\label{ss-sec:galex}

Our UV imaging comes from \textit{GALEX} near-UV (NUV) band. We download all \textit{GALEX} coadded images\footnote{\url{http://galex.stsci.edu/data}.} from \textit{GALEX} DR6$+$7 \citep{Bianchi2017} covering a region around each of our galaxies and mosaic a cutout sub-region around each source using \textsc{swarp} \citep{Bertin2002}. \textit{GALEX} does not have complete sky coverage in the WALLABY footprint, resulting in 122 (133) galaxies with (without) H\,\textsc{i} detections having \textit{GALEX} data. The majority of galaxies were only observed in the shallow \textit{GALEX} All-sky Imaging Survey (AIS), with a handful of galaxies observed in the deeper Medium Imaging Survey (MIS) and Nearby Galaxy Survey (NGS).

In a similar fashion to our PanSTARRS images, we create a segmentation map using the \textsc{photutils} task \textsc{segmentation}, which we use to mask objects other than the target galaxy within the image. We then determine and subtract the background by measuring the mean surface brightness in an annulus around the galaxy that does not contain any UV sources. We measure the background in five annuli of width 0.1 times the $r$-band radius convolved with the WISE W4 PSF (11.99\,arcsec). We convolve with the WISE W4 PSF as this band has the lowest resolution of the bands used for deriving SFRs (i.e.\ 4.9, 7.36 and 11.99\,arcsec for NUV, W3 and W4, respectively). The mean background is taken as the average of these five annuli. We measure the total flux in ADU (image units) in a series of apertures defined by the centre, position angle and inclination angle of the $r$-band aperture and radii scaled by $n\times0.1$ times the convolved $r$-band radius, where $n$ is the number of apertures, using the \textsc{photutils} task \textsc{aperture\_photometry}. The conversion from \textit{GALEX} image units ($y$) to magnitudes is
\begin{equation}
	m/\mathrm{mag} = 20.08 - 2.5\log(y),
	\label{equ:adu_to_mag_galex}
\end{equation}
where 20.08 is the zero point magnitude in the NUV-band. The \textsc{aperture\_photometry} task estimates the error, $\sigma_y$, in the measured aperture summed ADU based on the pixel error. We estimate the pixel error as the standard deviation of an annulus around the source aperture with the same position and inclination angles and the size defined by $a_{\rm{in}}=1.5a$, $a_{\rm{out}}=2.5a$, $b_{\rm{in}}=1.5b$ and $b_{\rm{out}}=2.5b$, where $a,b$ are the aperture major and minor axes. The source signal to noise ratio (SNR) is then $\rm{SNR}=\rm{ADU}_{\rm{aperture}}/\rm{ADU}_{\rm{error}}$. We consider NUV magnitudes with $\rm{SNR}<5$ to be upper limits. From the total magnitude in each aperture we get the total magnitude curve of growth for each galaxy. 

Using the curve of growth, we measure total asymptotic magnitudes \citep[e.g.][]{MunozMateos2015}. We perform a linear least squares fit to the total magnitude vs derivative of the curve of growth, $\Delta m$, for apertures where $\Delta mag<0.05$ (i.e.\ there is little variation in the curve of growth and it is approximately flat). The total asymptotic magnitude is then found by extrapolating this linear fit to $\Delta m_{\rm{fit}}=0.0$. We correct \textit{GALEX} NUV magnitudes for Galactic extinction following the same method as the PanSTARRS $g$- and $r$-bands, with $R_V=8.2$ from \cite{Wyder2007} for the NUV (Section~\ref{ss-sec:stellar_mass}).

\subsubsection{WISE}
\label{ss-sec:wise}

For our IR imaging, we use unWISE \citep{Lang2014,Meisner2017} coadds from WISE, which removes the blurring present in the original WISE coadded images. We follow the same method to create cutout sub-region images from unWISE as \textit{GALEX} from coadded intensity images.\footnote{W1 (\url{http://unwise.me/data/neo6/unwise-coadds/fulldepth}) and W3 and W4 (\url{http://unwise.me/data/allwise/unwise-coadds/fulldepth}).}

For the WISE bands (W1, W3, W4), we apply the mask created from the $r$-band image to mask other objects in the WISE band images. The resolution of WISE is lower than that of PanSTARRS (1.5\,arcsec) and the WISE PSF varies across the WISE bands (W1, W3, W4: 6.08, 7.36, 11.99\,arcsec, respectively). Hence we convolved the PanSTARRS aperture with the corresponding WISE band PSF prior to measuring the total magnitude within the aperture using the \textsc{photutils} task \textsc{aperture\_photometry}. The unWISE images are all set to the same zero point magnitude (22.5) and the image units ($y$) are converted to magnitudes using
\begin{equation}
	m/\mathrm{mag} = 22.5 - 2.5\log(y).
	\label{equ:adu_to_mag_unwise}
\end{equation}
We estimate the uncertainty in the WISE band magnitude in the same way as for the \textit{GALEX} NUV band and consider WISE band magnitudes with $\rm{SNR}<5$ to be upper limits.

\subsubsection{Star Formation Rate}
\label{ss-sec:star_formation_rate}

We calculate total SFRs for our galaxies by combining the contributions derived from \textit{GALEX} NUV and WISE mid-infrared (MIR) magnitudes following \cite{Janowiecki2017}. Both \textit{GALEX} and WISE have published all-sky source photometry catalogues, whose automated photometry measurements are optimised for point sources and tend to underestimate the total magnitude \citep[e.g.\ for WISE see][]{Jarrett2013}. As our sample includes all nearby, resolved galaxies, we measure magnitudes from image cutouts in the WISE and \textit{GALEX} photometric bands to ensure we recover all the source flux. We also use the derivative of the curve of growth to define the radius of the NUV disc emission (i.e.\ the size of the disc of recent star formation).

The UV SFR is derived from the \textit{GALEX} NUV band luminosity, $L_{\rm{NUV}}$, following \cite{Schiminovich2007} as,
\begin{equation}
	\mathrm{SFR_{NUV}}/(\mathrm{M}_{\odot}\mathrm{yr}^{-1}) = 10^{-28.165} L_{\mathrm{NUV}}/(\mathrm{erg\,s}^{-1}\,\rm{Hz}^{-1}).
	\label{equ:sfr_nuv}
\end{equation}
The MIR SFR comes from the WISE W4 band luminosity, $L_{\rm{W4}}$. If a galaxy is undetected in W4, then the W3 band luminosity, $L_{\rm{W3}}$, is used instead. The W3 and W4 WISE bands contain contamination from older stellar populations and require a correction by subtracting a fraction of the W1 band luminosity, $L_{\rm{W1}}$ \citep{Ciesla2014}. The W3 and W4 derived SFRs are given by \citep{Jarrett2013}
\begin{equation}
	\mathrm{SFR_{W3}}/(\mathrm{M}_{\odot}\mathrm{yr}^{-1}) = 4.91\times10^{-10} (L_{\mathrm{W3}} - 0.201 L_{\mathrm{W1}})/\mathrm{L}_{\odot}
	\label{equ:sfr_w3}
\end{equation}
and
\begin{equation}
	\mathrm{SFR_{W4}}/(\mathrm{M}_{\odot}\mathrm{yr}^{-1}) = 7.50\times10^{-10} (L_{\mathrm{W4}} - 0.044 L_{\mathrm{W1}})/\mathrm{L}_{\odot}.
	\label{equ:sfr_w4}
\end{equation}
We note that there are more recent WISE-based SFR calibrations from \cite{Cluver2017}, however we use the \cite{Jarrett2013} calibrations to enable comparison with results from xGASS \citep[][]{Janowiecki2017,Janowiecki2020} in Section~\ref{sec:results}. The total SFR is then
\begin{equation}
	\rm{SFR_{NUV+MIR}} = \rm{SFR_{NUV}} + \rm{SFR_{W4(3)}}.
	\label{equ:sfr_total}
\end{equation}
Our derived SFRs have uncertainties of $\lesssim0.1$\,dex.

Similarly to the $r$-band, we measure NUV-band disc diameters at an isophotal surface brightness of 28\,mag\,arcsec$^{-2}$. We also correct the NUV disc diameter for the PSF using Equation~\ref{equ:dhi_correction} adopting the NUV PSF of 4.9\,arcsec (FWHM).

\section{Results}
\label{sec:results}

\subsection{Sample Characterisation}
\label{s-sec:sample_characterisation}

\begin{figure}
	\includegraphics[width=\columnwidth]{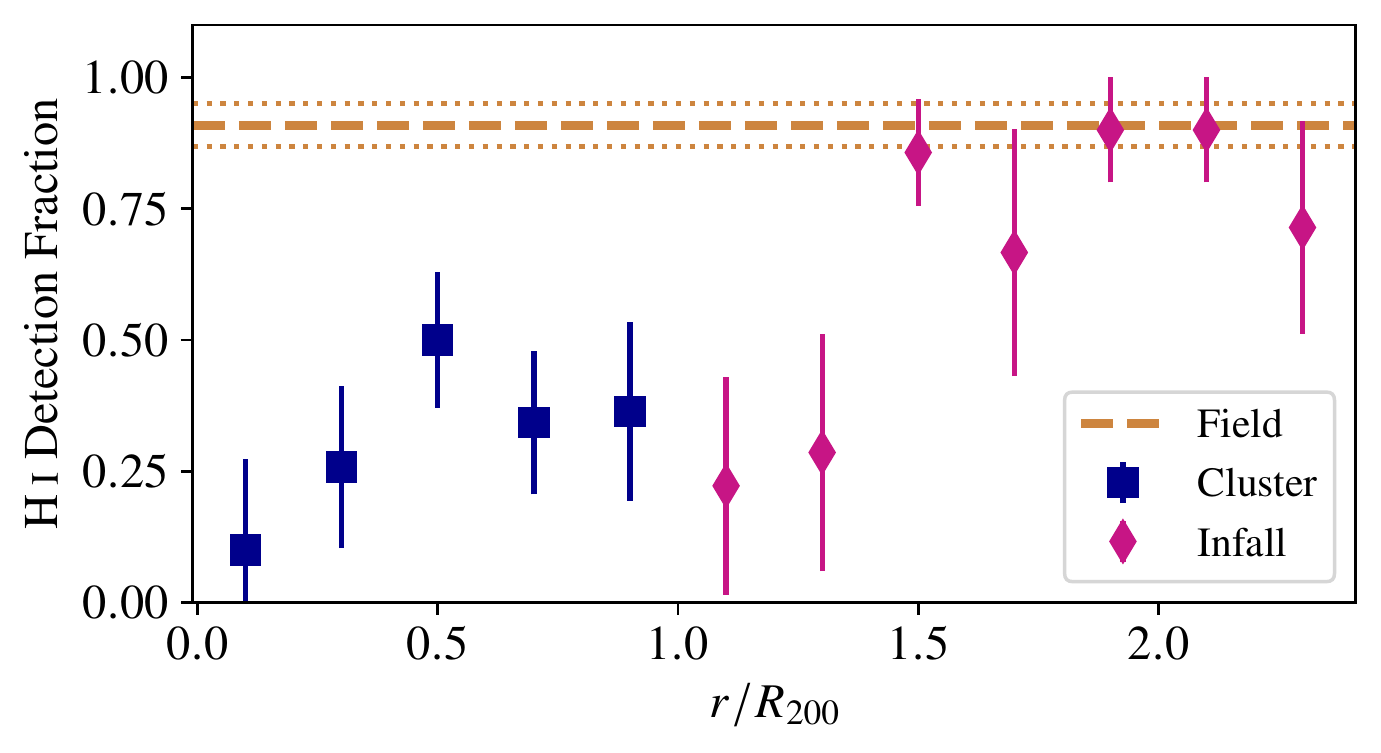}
    \caption{The fraction of galaxies detected in H\,\textsc{i} in radial bins of width $0.2R_{200}$ for the cluster and infall populations (blue squares and purple diamonds, respectively). The H\,\textsc{i} detected fraction of the field population is shown by the horizontal dashed orange line with the dotted lines indicating the uncertainty. The error bars are calculated assuming that the uncertainties follow a binomial distribution.}
    \label{fig:hi_detected_fraction}
\end{figure}

We begin by characterising the general properties of our sample of H\,\textsc{i} detections and non-detections. We quantify the fraction of galaxies detected in H\,\textsc{i} and where these galaxies lie with respect to the Hydra~I cluster centre. It is clear in Figures~\ref{fig:sky_pos} and \ref{fig:hydra_psd} that within $R_{200}$ the majority of galaxies are not detected in H\,\textsc{i}. In particular, only three galaxies detected in H\,\textsc{i} lie within the virialised region defined by \cite{Rhee2017} of Figure~\ref{fig:hydra_psd} (to the left of the red diagonal line). In Figure~\ref{fig:hi_detected_fraction}, we quantify the fraction of cluster (blue squares) and infall (purple diamonds) galaxies detected in H\,\textsc{i} as a function of projected distance from the cluster centre in radial bins of width $0.2R_{200}$, and compare this with the fraction of field (orange dashed line) galaxies detected in H\,\textsc{i}. Beyond $1.5R_{200}$, the fraction of galaxies detected in H\,\textsc{i} in the infall region is comparable to the H\,\textsc{i} detected fraction in the field ($\sim0.85$). The H\,\textsc{i} detected fraction drops sharply to $\sim0.3$ at $\sim1.5R_{200}$ and then decreases to $\sim0.2$ within the central $\sim0.2R_{200}$. 

\begin{figure}
	\includegraphics[width=\columnwidth]{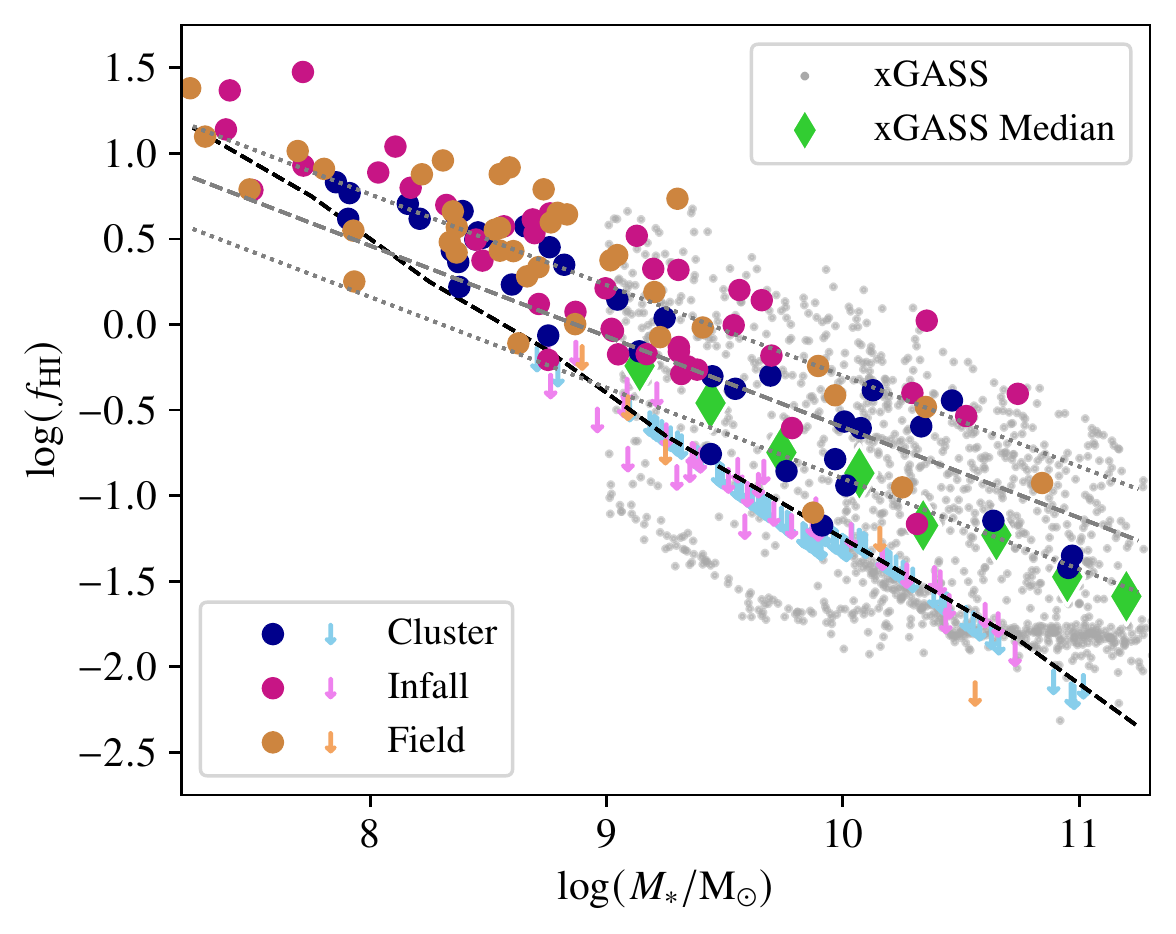}
    \caption{H\,\textsc{i} gas fraction ($f_{\rm{HI}}=M_{\rm{HI}}/M_{*}$) vs stellar mass ($M_{*}$) of galaxies in the Hydra field both with and without H\,\textsc{i} detections (filled circles and downward arrows, respectively). The cluster, infall and field populations are the blue, purple and orange symbols, respectively. The grey points are the xGASS sample, and the filled green diamonds show the median xGASS relation \citep{Catinella2018}. The dashed grey line is the gas fraction main sequence defined by \citet{Janowiecki2020} using star-forming galaxies in xGASS with the dotted grey lines indicating a scatter of 0.3\,dex above and below the main sequence. The dashed black line indicates the gas fraction sensitivity limit at the distance of the Hydra~I cluster (61\,Mpc).}
    \label{fig:hifrac_mstar}
\end{figure}

We compare our H\,\textsc{i} detections to the xGASS H\,\textsc{i} gas fraction ($f_{\rm{HI}}=M_{\rm{HI}}/M_{*}$) vs. stellar mass scaling relation from \cite{Catinella2018} in Figure~\ref{fig:hifrac_mstar}. xGASS is a stellar-mass selected ($M_*>10^9\,\rm{M}_{\odot}$), gas-fraction limited H\,\textsc{i} survey of $\sim1\,200$ galaxies with $0.02<z<0.05$, which is representative of the H\,\textsc{i} properties of galaxies in the local Universe. Around 50\% of the galaxies detected by WALLABY lie below the stellar-mass selection limit of xGASS ($M_*<10^9\,\rm{M}_{\odot}$). The H\,\textsc{i} non-detections identified from 6dFGS are predominantly galaxies in the same stellar-mass range probed by xGASS. As expected for a blind H\,\textsc{i} survey, WALLABY detects gas-rich galaxies (i.e.\ on or above the xGASS median relation from \citeauthor{Catinella2018} \citeyear{Catinella2018} for galaxies with $M_*>10^9\,\rm{M}_{\odot}$ and when extrapolated to lower stellar masses). Our H\,\textsc{i}-detected populations also tend to be more gas-rich compared to the gas fraction scaling relation defined by \cite{Janowiecki2020} using star-forming galaxies from xGASS (dashed grey line). We do not find any environmental dependence on the gas fractions of the H\,\textsc{i}-detected cluster, infall or field populations. H\,\textsc{i} stacking is required to probe the gas-poor regime at the Hydra~I cluster distance of 61\,Mpc.

\begin{figure}
	\includegraphics[width=\columnwidth]{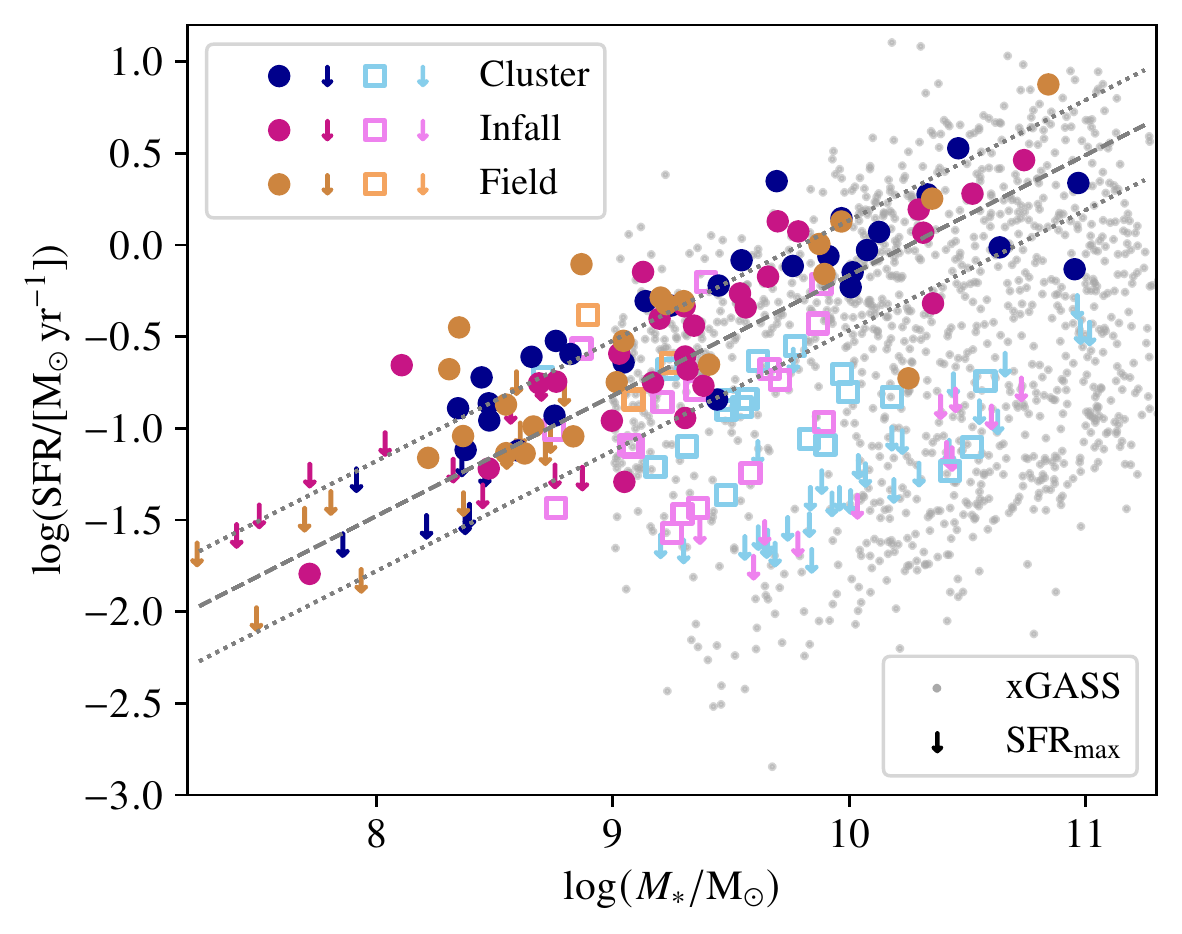}
    \caption{Star formation rate (SFR) vs stellar mass ($M_{*}$) for galaxies in the Hydra field both with and without H\,\textsc{i} detections. The colours are the same as in Figure~\ref{fig:hifrac_mstar}. The H\,\textsc{i} detections and non-detections with measures SFRs are shown by filled circles and unfilled squares, respectively. Upper-limit SFRs (SFR$_{\rm{max}}$) are indicated by downward arrows. The grey points are the xGASS sample. The star forming main sequence defined by \citet{Janowiecki2017} using xGASS is shown by the dashed black line with the dotted black lines indicating $\pm0.3$\,dex above and below the main sequence.}
    \label{fig:wallaby_sfms}
\end{figure}

We plot our cluster, infall and field populations of H\,\textsc{i} detections and non-detections in the star formation rate vs. stellar mass plane in Figure~\ref{fig:wallaby_sfms} and compare with the star forming main sequence defined using the xGASS sample by \cite{Janowiecki2017}. We can directly compare our galaxies to the xGASS star forming main sequence as we have derived SFRs in an identical way to \cite{Janowiecki2017}. All three H\,\textsc{i}-detected populations follow the xGASS-defined main sequence, while the majority of the cluster and infall H\,\textsc{i} non-detections lie below the main sequence in the region that contains passive, quenched galaxies. We investigate and quantify the distribution of our galaxy populations around the star forming main sequence below in Section~\ref{ss-sec:sfms_offset}.

\begin{figure}
	\includegraphics[width=\columnwidth]{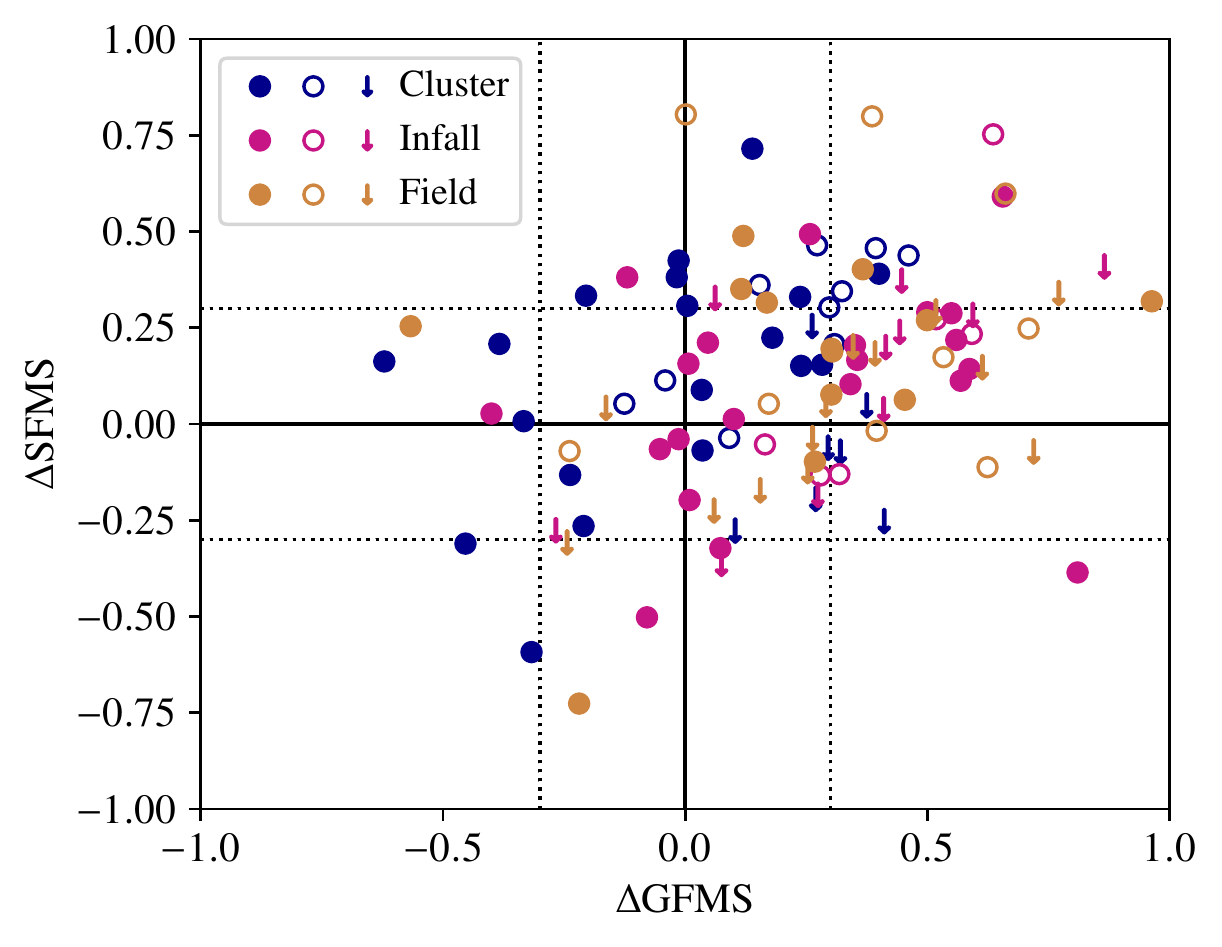}
\caption{Offset from the star forming main sequence ($\Delta \rm{SFMS}$) vs offset from the gas fraction main sequence ($\Delta \rm{GFMS}$) for galaxies detected in H\,\textsc{i}. The colours are the same as in Figure~\ref{fig:hifrac_mstar}. Upper-limit SFRs are indicated by downward arrows and are all low-mass ($M_*<10^9\,\rm{M}_{\odot}$). Low-mass galaxies with measured SFRs are indicated by unfilled circles. The star forming and gas fraction main sequences in Figures~\ref{fig:hifrac_mstar} and \ref{fig:wallaby_sfms} are shown by the solid black lines with the dotted black lines indicating $\pm0.3$\,dex above and below the main sequences.}
    \label{fig:sfms_gfms}
\end{figure}

In Figure~\ref{fig:sfms_gfms}, we look at the offset of galaxies above and below the star forming main sequence, $\Delta \rm{SFMS}$, and gas fraction main sequence, $\Delta \rm{GFMS}$. The offset is defined as the vertical displacement (i.e.\ at fixed stellar mass) of each galaxy above or below the SFMS or GFMS. Unsurprisingly, we find that galaxies are concentrated in the top right quadrant (i.e.\ positive  $\Delta \rm{SFMS}$ and $\Delta \rm{GFMS}$). This indicates that galaxies which are more gas-rich are also more star forming. We also find possible evidence for gas removal that is not yet affecting star formation (i.e.\ moving horizontally across Figure~\ref{fig:sfms_gfms}) between the infall and cluster populations where the cluster points (blue) appear to the preferentially located to the left of the infall (purple) points, although we note that the majority of galaxies lie within the $1\sigma$ scatter of the main sequences. Additionally, the GFMS below $M_*=10^9\,\rm{M}_{\odot}$ is just an extrapolation and no strong conclusions should be drawn for galaxies with $M_*<10^9\,\rm{M}_{\odot}$. However, we find that the median $\Delta \rm{GFMS}$ of high mass ($M_*\geq10^9\,\rm{M}_{\odot}$) cluster galaxies is lower than that of the infall and field, but there is no corresponding decrease in the median $\Delta \rm{SFMS}$ (Section~\ref{ss-sec:sfms_offset}).

\subsection{Probing the Environment}
\label{s-sec:probe_environment}

\begin{figure*}
	\includegraphics[width=15cm]{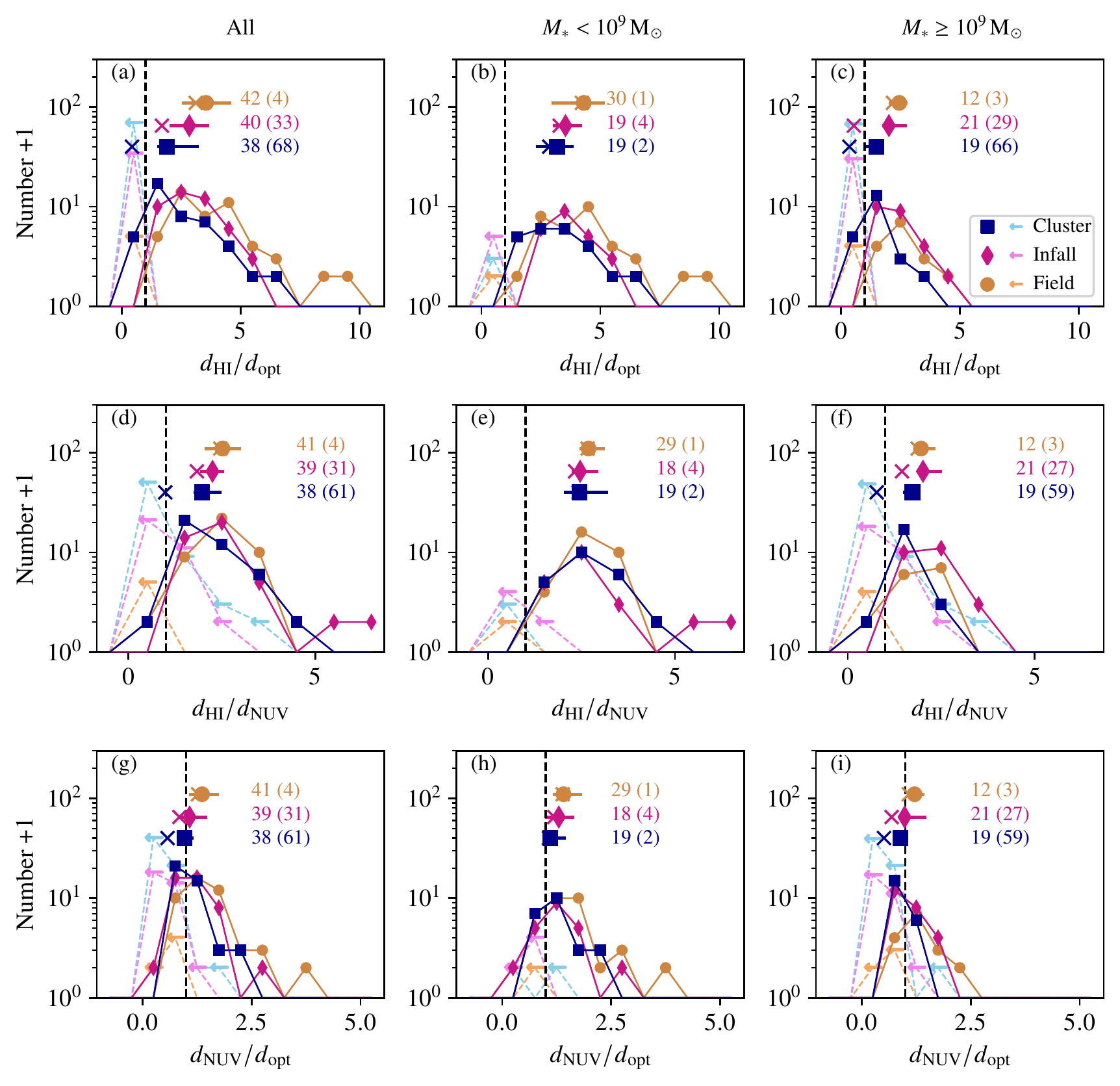}
    \caption{Histograms of the H\,\textsc{i} to optical $r$-band diameter ratio, $d_{\rm{HI}}/d_{\rm{opt}}$, the H\,\textsc{i} to NUV disc diameter ratio, $d_{\rm{HI}}/d_{\rm{NUV}}$ and the NUV to optical $r$-band disc diameter ratio, $d_{\rm{NUV}}/d_{\rm{opt}}$ (top, centre and bottom rows, respectively) for the three galaxy populations: cluster, infall and field (blue square, purple diamond and orange circle, respectively). The solid dark lines/symbols show the distributions of H\,\textsc{i}-detected galaxies. The dashed light lines/left-facing arrows show the distributions for H\,\textsc{i} non-detections. The large solid points and horizontal error bars indicate the median and $25^{\rm{th}}$ and $75^{\rm{th}}$ percentiles for the-H\,\textsc{i} detected distributions (tabulated in Table~\ref{table:sample_medians}). The crosses indicate the medians including the H\,\textsc{i} non-detections. The numbers to the left of the medians are the total number of H\,\textsc{i} detections (non-detections) in each population. The vertical dashed lines indicate a diameter ratio of 1. The left column includes all galaxies. The centre and right columns show low- ($M_*<10^9\,\rm{M}_{\odot}$) and high-mass ($M_*\geq10^9\,\rm{M}_{\odot}$) subsamples, respectively. To ease the differentiation between bins with 0 or 1 count, we have plotted the total in each bin as one more than the number of galaxies (i.e.\ $10^0$ means the bin contains no galaxies).}
    \label{fig:histograms_diameter}
\end{figure*}

\begin{figure*}
	\includegraphics[width=15cm]{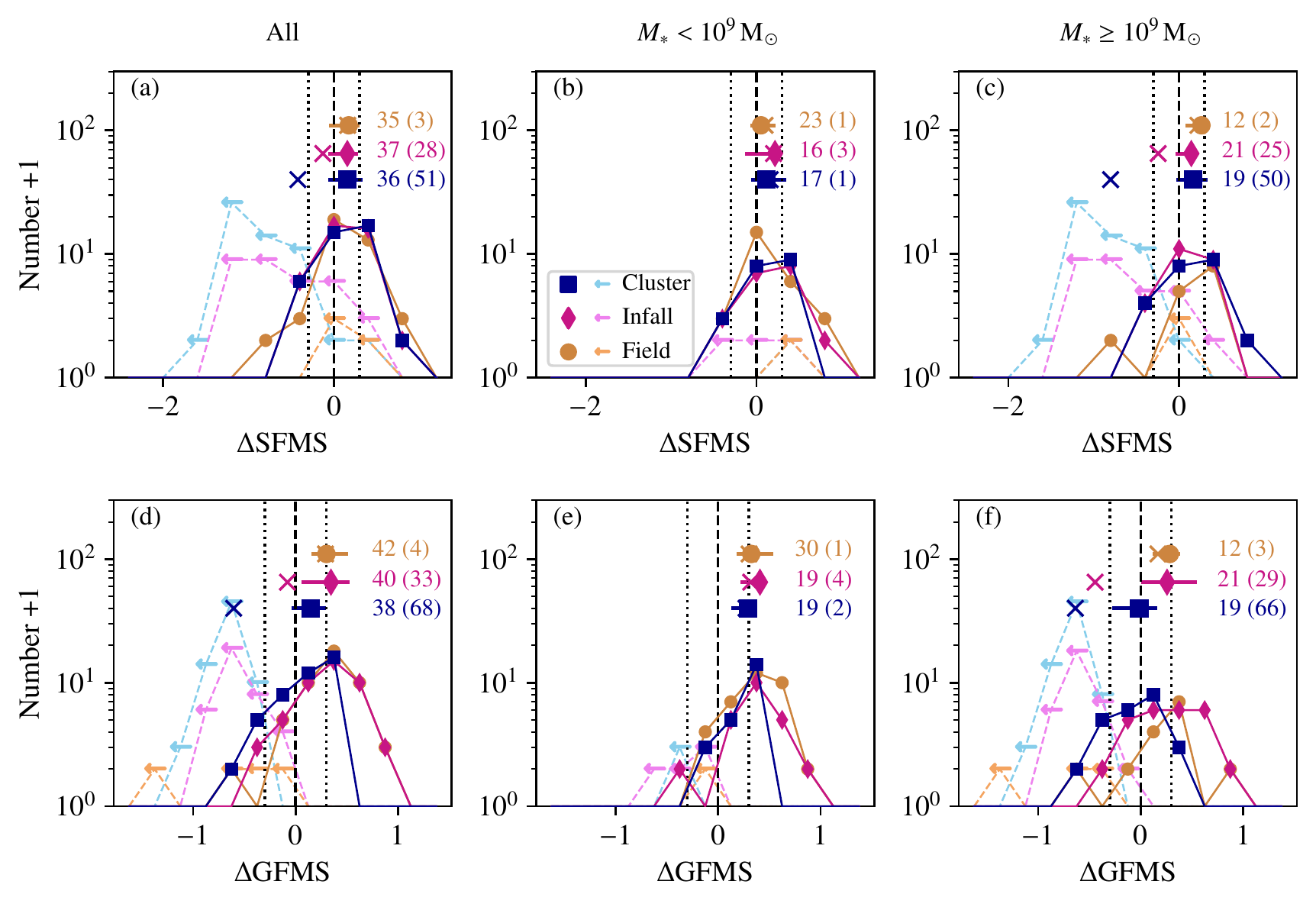}
    \caption{Similar to Figure~\ref{fig:histograms_diameter} showing histograms of the distance from the star forming main sequence, $\Delta \rm{SFMS}$, and the distance from the gas fraction main sequence, $\Delta \rm{GFMS}$ (top and bottom rows, respectively). The medians and $25^{\rm{th}}$ and $75^{\rm{th}}$ percentiles are tabulated in Table~\ref{table:sample_medians_ms}. The xGASS star forming and gas fraction main sequences (vertical lines in the top and bottom panels, respectively) as in Figure~\ref{fig:sfms_gfms}.}
    \label{fig:histograms_ms}
\end{figure*}

We investigate the influence of the environment on gas removal and star formation of galaxies falling into Hydra~I using the ratios between H\,\textsc{i}, $r$-band and NUV disc diameters (Figure~\ref{fig:histograms_diameter}) and the offsets from the star forming and gas fraction main sequences (Figure~\ref{fig:histograms_ms}). In these figures, we show the distributions of each measured property for our H\,\textsc{i} detected cluster, infall and field populations (blue square, purple diamond and orange circle, respectively). The rows of Figure~\ref{fig:histograms_diameter} show, from top to bottom, histograms of the distributions of H\,\textsc{i} to optical $r$-band disc diameter ratio ($d_{\rm{HI}}/d_{\rm{opt}}$), the H\,\textsc{i} to NUV disc diameter ratio ($d_{\rm{HI}}/d_{\rm{NUV}}$) and the NUV to optical $r$-band disc diameter ratio ($d_{\rm{NUV}}/d_{\rm{opt}}$). The top and bottom rows of Figure~\ref{fig:histograms_ms} show histograms of the offset from the star forming main sequence ($\Delta \rm{SFMS}$) and the offset from the gas fraction main sequence ($\Delta \rm{GFMS}$). Above the histograms, we plot the median and 25$^{\rm{th}}$ and 75$^{\rm{th}}$ percentiles of each distribution. The H\,\textsc{i} non-detections are shown using light-coloured downward triangles and dashed lines with the median of the combined H\,\textsc{i} detections and non-detections shown by the crosses. 

The left columns of Figures~\ref{fig:histograms_diameter} and \ref{fig:histograms_ms} include all galaxies in each population. Galaxy properties show a dependence on stellar mass \citep[e.g. gas fraction and star formation rate; see Figures~\ref{fig:hifrac_mstar} and \ref{fig:wallaby_sfms} of this work and][]{Catinella2018}. To check if trends present in the full sample are driven by stellar mass, we subdivide the galaxy populations into low- and high-stellar-mass subsamples ($M_*<10^9$ and $\geq10^9\,\rm{M}_{\odot}$, i.e.\ boundary between dwarf and giant galaxies) in the centre and right columns. Splitting the sample at $M_*=10^9\,\rm{M}_{\odot}$ also produces a fairly even number of H\,\textsc{i}-detected galaxies in each subsample. Although we note that the majority of the H\,\textsc{i} non-detections are in the high-mass subsample. Moreover, not all quantities can be measured for all galaxies (e.g.\ some galaxies with PanSTARRS imaging are missing \textit{GALEX} coverage, while others with \textit{GALEX} imaging have artefacts in the PanSTARRS images) which results in different galaxy totals. Our results do not change if we limit our sample to galaxies measured in all wavelengths and we choose to use all galaxies for which we can measure each quantity to maximise our statistics. We discuss Figures~\ref{fig:histograms_diameter} and \ref{fig:histograms_ms} in more detail below.

\begin{table*}
	\centering
	\caption{Median H\,\textsc{i} to optical $r$-band diameter ratio ($d_{\rm{HI}}/d_{\rm{opt}}$), H\,\textsc{i} to NUV diameter ratio ($d_{\rm{HI}}/d_{\rm{NUV}}$) and NUV to optical diameter ratio ($d_{\rm{NUV}}/d_{\rm{opt}}$) for the cluster, infall and field populations in Figure~\ref{fig:histograms_diameter}. For the H\,\textsc{i}-detected samples, the 25$^{\rm{th}}$ and 75$^{\rm{th}}$ percentiles are the median $\pm$ the superscript and subscripts, respectively.}
	\label{table:sample_medians}
	\begin{tabular}{lccccccccr}
	    \hline
		& \multicolumn{3}{c}{$d_{\rm{HI}}/d_{\rm{opt}}$} & \multicolumn{3}{c}{$d_{\rm{HI}}/d_{\rm{NUV}}$} & \multicolumn{3}{c}{$d_{\rm{NUV}}/d_{\rm{opt}}$} \\
		Sample                         & Cluster   & Infall    & Field      & Cluster   & Infall    & Field     & Cluster   & Infall    & Field \\
		\hline
		\vspace{1.5pt}
		& \multicolumn{9}{c}{H\,\textsc{i} Detections} \\
		\vspace{1.5pt}
		All                            & 1.9$^{+1.3}_{-0.4}$ & 2.8$^{+0.8}_{-0.8}$ & 3.5$^{+1.1}_{-1.0}$ & 2.0$^{+0.5}_{-0.2}$ & 2.2$^{+0.3}_{-0.3}$ & 2.5$^{+0.5}_{-0.5}$ & 1.0$^{+0.2}_{-0.2}$ & 1.1$^{+0.4}_{-0.2}$ & 1.4$^{+0.4}_{-0.3}$   \\
		\vspace{1.5pt}
		$M_*<10^9\,\rm{M}_{\odot}$     & 3.2$^{+0.7}_{-0.9}$ & 3.5$^{+0.7}_{-0.5}$ & 4.3$^{+0.9}_{-1.4}$ & 2.4$^{+0.8}_{-0.4}$ & 2.5$^{+0.5}_{-0.2}$ & 2.7$^{+0.4}_{-0.2}$ & 1.1$^{+0.4}_{-0.2}$ & 1.3$^{+0.4}_{-0.3}$ & 1.4$^{+0.4}_{-0.2}$   \\
		$M_*\geq10^9\,\rm{M}_{\odot}$  & 1.5$^{+0.2}_{-0.4}$ & 2.0$^{+0.8}_{-0.3}$ & 2.5$^{+0.4}_{-0.4}$ & 1.7$^{+0.2}_{-0.2}$ & 2.0$^{+0.5}_{-0.1}$ & 2.0$^{+0.4}_{-0.2}$ & 0.9$^{+0.1}_{-0.1}$ & 1.0$^{+0.5}_{-0.1}$ & 1.2$^{+0.2}_{-0.2}$   \\ \\
		
		\vspace{1.5pt}
		& \multicolumn{9}{c}{H\,\textsc{i} Detections and Non-Detections} \\
		
		All                            & 0.4    & 1.7   & 3.1   & 1.0   & 1.8   & 2.5   & 0.6   & 0.8   & 1.3   \\
		$M_*<10^9\,\rm{M}_{\odot}$     & 2.9    & 3.3   & 4.2   & 2.4   & 2.3   & 2.7   & 1.1   & 1.2   & 1.4    \\
		$M_*\geq10^9\,\rm{M}_{\odot}$  & 0.4    & 0.5   & 2.1   & 0.8   & 1.5   & 1.9   & 0.5   & 0.7   & 1.1   \\ \\
		
		\vspace{1.5pt}
		& \multicolumn{9}{c}{H\,\textsc{i} Detections ($d_{\rm{HI}}>45$\,arcsec)} \\
		\vspace{1.5pt}
		All                            & 1.9$^{+1.4}_{-0.5}$ & 2.6$^{+1.1}_{-0.8}$ & 3.5$^{+1.2}_{-1.0}$ & 2.0$^{+0.5}_{-0.2}$ & 2.3$^{+0.3}_{-0.3}$ & 2.5$^{+0.5}_{-0.5}$ & --- & --- & ---   \\
		\vspace{1.5pt}
		$M_*<10^9\,\rm{M}_{\odot}$     & 3.2$^{+0.8}_{-0.7}$ & 3.6$^{+0.8}_{-0.6}$ & 4.2$^{+1.1}_{-1.3}$ & 2.4$^{+0.8}_{-0.4}$ & 2.5$^{+0.4}_{-0.2}$ & 2.7$^{+0.5}_{-0.2}$ & --- & --- & ---   \\
		$M_*\geq10^9\,\rm{M}_{\odot}$  & 1.5$^{+0.2}_{-0.4}$ & 2.0$^{+0.7}_{-0.3}$ & 2.5$^{+0.4}_{-0.4}$ & 1.7$^{+0.2}_{-0.2}$ & 2.0$^{+0.5}_{-0.2}$ & 2.0$^{+0.3}_{-0.2}$ & --- & --- & ---   \\
		
		\hline
	\end{tabular}
\end{table*}

\begin{table*}
	\centering
	\caption{Median offset from the star forming main sequence ($\Delta \rm{SFMS}$) and gas fraction main sequence ($\Delta \rm{GFMS}$) for the cluster, infall and field populations in Figure~\ref{fig:histograms_ms}. For the H\,\textsc{i}-detected samples, the 25$^{\rm{th}}$ and 75$^{\rm{th}}$ percentiles are the median $\pm$ the superscript and subscripts, respectively.}
	\label{table:sample_medians_ms}
	\begin{tabular}{lcccccr}
	    \hline
		& \multicolumn{3}{c}{$\Delta \rm{SFMS}$} & \multicolumn{3}{c}{$\Delta \rm{GFMS}$} \\
		Sample                         & Cluster   & Infall    & Field      & Cluster   & Infall    & Field  \\
		\hline
		\vspace{1.5pt}
		& \multicolumn{6}{c}{H\,\textsc{i} Detections} \\
		\vspace{1.5pt}
		All                            & 0.2$^{+0.2}_{-0.2}$ & 0.2$^{+0.1}_{-0.2}$ & 0.2$^{+0.1}_{-0.2}$ & 0.1$^{+0.2}_{-0.2}$ & 0.3$^{+0.2}_{-0.3}$ & 0.3$^{+0.2}_{-0.1}$  \\
		\vspace{1.5pt}
		$M_*<10^9\,\rm{M}_{\odot}$     & 0.1$^{+0.2}_{-0.2}$ & 0.2$^{+0.1}_{-0.3}$ & 0.1$^{+0.2}_{-0.1}$ & 0.3$^{+0.0}_{-0.2}$ & 0.4$^{+0.1}_{-0.2}$ & 0.3$^{+0.2}_{-0.1}$  \\
		$M_*\geq10^9\,\rm{M}_{\odot}$  & 0.2$^{+0.2}_{-0.2}$ & 0.1$^{+0.1}_{-0.2}$ & 0.3$^{+0.1}_{-0.2}$ & 0.0$^{+0.2}_{-0.3}$ & 0.3$^{+0.3}_{-0.3}$ & 0.3$^{+0.1}_{-0.2}$. \\ \\
		
		\vspace{1.5pt}
		& \multicolumn{6}{c}{H\,\textsc{i} Detections and Non-Detections} \\
		
		All                            & $-0.4$    & $-0.1$    & 0.2     & $-0.6$   & $-0.1$   & 0.3   \\
		$M_*<10^9\,\rm{M}_{\odot}$     & 0.2       & 0.2       & 0.1     & 0.3   & 0.3   & 0.3  \\
		$M_*\geq10^9\,\rm{M}_{\odot}$  & $-0.8$    & $-0.2$    & 0.2     & $-0.6$   & $-0.4$   & 0.2  \\
		\hline
	\end{tabular}
\end{table*}

\begin{figure*}
	\includegraphics[width=\textwidth]{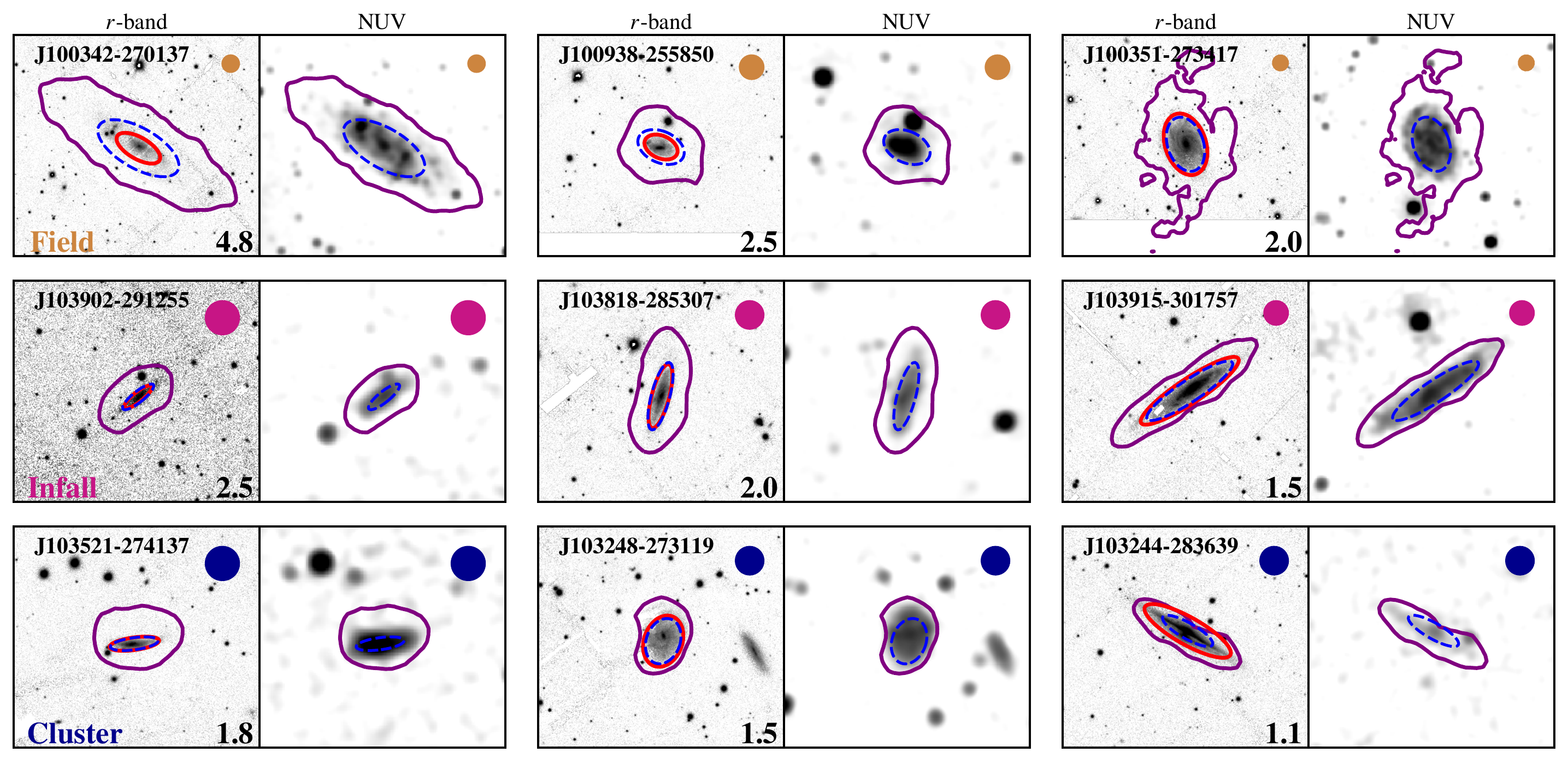}
    \caption{Example cluster, infall and field galaxies (top, centre and bottom rows, respectively). The three galaxies from each population are taken from across the sample stellar mass range and have similar stellar masses of $M_*\sim10^{8.7}$, $10^{9.5}$ and $10^{10.5}\,\rm{M}_{\odot}$ (left, centre and right columns, respectively). The PanSTARRS $r$-band and \textit{GALEX} NUV-band images for each galaxy are shown, with overlaid ellipses (solid red and dashed blue, respectively) showing the measured sizes in each band. The purple contour shows the inclination-corrected H\,\textsc{i} surface density of $1\,\rm{M}_{\odot}\,\rm{pc}^{-2}$. The filled circle in the top right corner of each panel shows the WALLABY synthesised beam, which is coloured according to its parent population: cluster (blue), infall (purple), field (orange). The number in the bottom right of the $r$-band panels is the measured $d_{\rm{HI}}/d_{\rm{opt}}$ of each galaxy. For added clarity, the NUV images are smoothed using a $3\times3$ pixel kernel. The NUV disc diameters are measured using the original, unsmoothed data.}
    \label{fig:galaxy_plate}
\end{figure*}

\subsubsection{Atomic Gas Disc Extent}
\label{ss-sec:size_ratio}

The H\,\textsc{i} to isophotal 23.5\,mag\,arcsec$^{-2}$ $r$-band disc diameter ratio, $d_{\rm{HI}}/d_{\rm{opt}}$, quantifies the relative size of the H\,\textsc{i} disc for a uniformly defined optical disc surface brightness and can be used to probe gas stripping. This parameter is most sensitive to gas removal in the outer disc (i.e.\ where environmental effects will be strongest and first felt by a galaxy). We find a clear decreasing trend in the median $d_{\rm{HI}}/d_{\rm{opt}}$ from the field, $d_{\rm{HI}}/d_{\rm{opt}}=3.5$, to the cluster, $d_{\rm{HI}}/d_{\rm{opt}}=2.0$ ($\sim43\%$ smaller $d_{\rm{HI}}/d_{\rm{opt}}$ for the cluster population, panel a of Figure~\ref{fig:histograms_diameter}). We assess the significance of difference in the $d_{\rm{HI}}/d_{\rm{opt}}$ distributions using the Kolmogorov–Smirnov (KS) test, where the difference between the distributions is statistically significant for $p$-values $<0.05$. For $p$-values $\geq0.05$, we reject the hypothesis that the distributions are different. We find that the cluster population is statistically different from the infall and field ($p$-values of 0.0061 and 0.0001), while the difference in the infall and field distributions is not statistically significant ($p$-value of 0.06). The reduction in the diameter ratio found in denser environments suggests that the environment is responsible for removing gas from the infall of the galaxies, thereby shrinking their H\,\textsc{i} discs. This trend is strengthened if H\,\textsc{i} non-detections with upper limits for $d_{\rm{HI}}/d_{\rm{opt}}$ are included in the median (the cluster median $d_{\rm{HI}}/d_{\rm{opt}}$ is 90\% smaller than that of the field). The upper limits from galaxies not detected in H\,\textsc{i} with WALLABY are all truncated within the optical disc (i.e.\ $d_{\rm{HI}}/d_{\rm{opt}}<1$, see Section~\ref{ss-sec:hi_diameter} for our method of estimating H\,\textsc{i} non-detection $d_{\rm{HI}}$ upper limits). Hence there are likely populations of cluster and infall galaxies with H\,\textsc{i} discs smaller than their optical discs, similar to those observed in the Virgo cluster \citep{Chung2009,Yoon2017}, which are below the WALLABY mass detection limit at the distance of Hydra ($M_{\rm{HI}}\sim10^{8.7}\,\rm{M}_{\odot}$ at 61\,Mpc). 

The median decrease in $d_{\rm{HI}}/d_{\rm{opt}}$ from the field to the cluster is not mass-dependent, as the same trend is present in low- and high-mass ($M_*<10^9$ and $>10^9\,\rm{M}_{\odot}$, respectively) subsamples (panels b and c of Figure~\ref{fig:histograms_diameter}). These subsamples also show that lower-mass galaxies tend to have larger, more extended H\,\textsc{i} gas reservoirs compared their optical discs, which is expected from the H\,\textsc{i} gas fraction vs stellar mass scaling relation, where gas fraction decreases with increasing stellar mass (Figure~\ref{fig:hifrac_mstar}). We tabulate the medians and 25$^{\rm{th}}$ and 75$^{\rm{th}}$ percentiles from Figure~\ref{fig:histograms_diameter} in Table~\ref{table:sample_medians}. Our results do not change if we exclude unresolved H\,\textsc{i} detections for which our measured $d_{\rm{HI}}$ are upper limits (see Section~\ref{ss-sec:hi_diameter}).

To illustrate the H\,\textsc{i} disc truncation observed as galaxies fall into Hydra~I, we show PanSTARRS $r$-band and \textit{GALEX} NUV band images for three cluster, infall and field galaxies in Figure~\ref{fig:galaxy_plate} (top, centre and bottom rows, respectively). The measured sizes of the $r$-band, NUV and H\,\textsc{i} discs are shown by the overlaid solid red and dashed blue ellipses and purple contour, respectively. The columns show a galaxy from each population with similar stellar masses of $M_*\sim10^{8.7}$, $10^{9.5}$ and $10^{10.5}\,\rm{M}_{\odot}$ (left, centre and right columns, respectively). We observe the H\,\textsc{i} disc truncation across all stellar masses (lower rows of Figure~\ref{fig:histograms_diameter}). We also see that H\,\textsc{i} discs are more extended compared to their optical discs in the low-mass galaxies than the high-mass galaxies. Although we note that due to the H\,\textsc{i} detection limit, it will be more difficult to detect H\,\textsc{i}-poor low-mass galaxies.

\subsubsection{Offset from the Star Forming and Gas Fraction Main Sequences}
\label{ss-sec:sfms_offset}

At a fixed stellar mass, galaxies significantly ($>0.3$\,dex) above (below) the star forming main sequence are forming more (less) stars than those that lie on the main sequence, and are referred to as star-bursting (quenched). Hence, a galaxy population's offset from the star forming main sequence at fixed stellar mass provides information on the population's average star formation relative to normal star-forming galaxies. Similarly, the offset of a galaxy population from the gas fraction main sequence for star-forming galaxies provides an indication of population's gas-richness.

Panel a of Figure~\ref{fig:histograms_ms} collapses the star formation rate vs stellar mass plot (Figure~\ref{fig:wallaby_sfms}) in stellar mass to show the distribution of the cluster, infall and field populations about the star forming main sequence, $\Delta \rm{SFMS}$, (vertical dashed line at 0) from xGASS. The three populations of H\,\textsc{i}-detected galaxies are centred on the star forming main sequence and are indistinguishable (KS test $p$-values between each population pair are $>0.5$). There is only a clear difference in median offset below the star forming main sequence for the cluster population once H\,\textsc{i} non-detections are included (KS test $p$-value of $<0.0001$ between the cluster and field populations). These quenched galaxies in the cluster and infall populations are responsible for the peaks in the H\,\textsc{i} non-detections at $\sim1.2$\,dex below the star forming main sequence (i.e.\ $\sim4\sigma$ below the star forming main sequence). The H\,\textsc{i} non-detections dominate the cluster population ($\sim66\%$) and lower the median offset to below the star forming main sequence by $\sim0.8$\,dex. The median of the infall population also decreases, but remains within the star forming main sequence scatter (non-detections only account for $\sim43\%$ of the infall population). The three H\,\textsc{i} non-detections in the field lie on the star forming main sequence and have stellar masses $<10^{9.3}\,\rm{M}_{\odot}$. We find no dependence on stellar mass if we split the populations into low and high stellar mass (panels b and c of Figure~\ref{fig:histograms_ms}). Similar to the $\Delta \rm{SFMS}$ behaviour for the three populations, we find no difference in the distributions of $\rm{NUV}-r$ colour for the cluster, infall and field galaxies.

We plot the distributions of $\Delta \rm{GFMS}$ for the cluster, infall and field galaxies in panel d of Figure~\ref{fig:histograms_ms}. As hinted at in Figure~\ref{fig:sfms_gfms}, we see that the cluster population median is offset by $\sim66\%$ to lower $\Delta \rm{GFMS}$ compared to the infall or field. Performing a KS test, we find that the difference between the cluster and infall/field populations is statistically significant ($p$-values of 0.003 and 0.024 with the infall and field, respectively). Although all medians are above $\Delta \rm{GFMS}=0$, which illustrates that we are most sensitive to gas-rich and gas-normal galaxies. There are no cluster galaxies with $\Delta \rm{GFMS}>0.5$ compared to 12 and 9 infall and field galaxies, respectively. This follows our findings of smaller $d_{\rm{HI}}/d_{\rm{opt}}$ moving from the field to the cluster (panel a of Figure~\ref{fig:histograms_diameter}) as the gas fraction will presumably be lower in galaxies with smaller $d_{\rm{HI}}/d_{\rm{opt}}$. Similar to the results for $d_{\rm{HI}}/d_{\rm{opt}}$ and $\Delta \rm{SFMS}$, including H\,\textsc{i} increases the difference in medians among the three populations by lowering the cluster median to $\sim2\sigma$ below $\Delta \rm{GFMS}=0$ and the infall median to $\Delta \rm{GFMS}\sim0.1$ due to large number of H\,\textsc{i} non-detections in these two populations. Unlike $d_{\rm{HI}}/d_{\rm{opt}}$ and $\Delta \rm{SFMS}$, the results for $\Delta \rm{GFMS}$ show a mass dependence. In the low-mass subsample (panel e of Figure~\ref{fig:histograms_ms}) there is no difference in the median offset. This is likely due to our extrapolation of the gas fraction main sequence below $M_*<10^9\,\rm{M}_{\odot}$. For the high-mass subsample (panel f of Figure~\ref{fig:histograms_ms}), the trend of decreasing $\Delta \rm{GFMS}$ from the field to the cluster remains with the cluster and infall medians 100\% and 33\% lower than the field. The significance of the difference between the cluster and field populations is unchanged. However, the difference between the cluster and infall populations becomes less significant (KS test $p$-value of 0.038).

\subsubsection{Near-UV Disc Extent}
\label{ss-sec:nuv_disc}

Although the H\,\textsc{i}-detected cluster population is not offset below the star forming main sequence, indicating no disc-wide quenching is occurring, there may be a reduction in the star formation in the outer disc (i.e.\ beyond the optical disc). This would indicate that gas is being predominantly stripped from the disc outskirts. NUV emission traces recently formed, young stars and we find that it extends beyond the optical $r$-band disc in a large fraction of galaxies detected in H\,\textsc{i} ($d_{\rm{NUV}}>d_{\rm{opt}}$, panel g of Figure~\ref{fig:histograms_diameter}, see also Figure~\ref{fig:galaxy_plate}). We note that our results for the NUV disc sizes are true for diameters measured to an isophotal surface brightness of 28\,mag\,arcsec$^{-2}$. With sufficiently deep observations, galaxies are found to possess NUV discs with comparable extents as in the H\,\textsc{i} \citep[e.g.][]{Meurer2018}.

We compare the H\,\textsc{i} to NUV ($d_{\rm{HI}}/d_{\rm{NUV}}$) and NUV to optical ($d_{\rm{NUV}}/d_{\rm{opt}}$) disc diameter ratios of the three galaxy populations in panels d and g of Figure~\ref{fig:histograms_diameter}, respectively. We find a small decrease in the median $d_{\rm{HI}}/d_{\rm{NUV}}$ and $d_{\rm{NUV}}/d_{\rm{opt}}$ moving from the field, through the infall and to the cluster, although these trends are more subtle than for the H\,\textsc{i} to optical diameter ratio (i.e.\ $\sim40\%$ smaller $d_{\rm{HI}}/d_{\rm{opt}}$ vs $\sim20\%$ smaller $d_{\rm{HI}}/d_{\rm{NUV}}$ and $\sim30\%$ smaller $d_{\rm{NUV}}/d_{\rm{opt}}$ between the cluster and field). Only the differences between the cluster and field $d_{\rm{HI}}/d_{\rm{NUV}}$ and $d_{\rm{NUV}}/d_{\rm{opt}}$ distributions are statistically significant according to a KS test ($p$-values of 0.0068 and 0.0004, respectively). Including H\,\textsc{i} non-detection upper limits has a similar effect on $d_{\rm{HI}}/d_{\rm{NUV}}$ as we find for $d_{\rm{HI}}/d_{\rm{opt}}$ (the cluster median becomes 80\% smaller than the field). This indicates two things: (1) that galaxies in the cluster population have smaller gas reservoirs than galaxies with similar NUV discs in the field and (2) that the cluster galaxies have on average smaller star-forming discs than the field population. The stronger trend in $d_{\rm{HI}}/d_{\rm{opt}}$ than $d_{\rm{NUV}}/d_{\rm{opt}}$ with environment also illustrates the sensitivity of H\,\textsc{i} for probing environmental effects. We find the same trends of decreasing median $d_{\rm{HI}}/d_{\rm{NUV}}$ and $d_{\rm{NUV}}/d_{\rm{opt}}$ in subsamples of low and high mass galaxies (panels e, f, h and i of Figure~\ref{fig:histograms_diameter}). The truncation of the NUV disc relative to the $r$-band disc is also illustrated in Figure~\ref{fig:galaxy_plate} (dashed blue and solid red ellipses, respectively). We note that we have smoothed the NUV images for clarity, but measure the NUV disc sizes from the original, unsmoothed data. This smoothing is the cause of the apparent underestimation of the NUV disc diameters.

\section{Discussion}
\label{sec:discussion}

\subsection{Gas Removal}
\label{s-sec:disc_trucation}

In agreement with the results of previous studies \citep[e.g.][]{Giovanelli1985,Solanes2001,Hess2013,Brown2017,Yoon2017}, we find clear evidence that the environment of Hydra~I is affecting the H\,\textsc{i} gas content and extent of galaxies in the cluster and infall populations. The fraction of galaxies detected in H\,\textsc{i} within $<1.5R_{200}$ of Hydra~I ($\sim0.3$) is significantly lower than that of infall galaxies at $>1.5R_{200}$ ($\sim0.8$). The H\,\textsc{i} detected fraction of infall galaxies at $>1.5R_{200}$ is also comparable to the fraction of galaxies detected in the field ($\sim0.9$). This indicates that environmental processes are likely responsible for the removal and/or depletion of H\,\textsc{i} to below the WALLABY detection limit of $M_{\rm{HI}}\sim10^{8.7}\,\rm{M}_{\odot}$ at projected distances of $<1.5R_{200}$ from Hydra~I. 

A significant population of galaxies ($\sim50\%$) in the VLA Imaging survey of Virgo galaxies in Atomic gas \citep[VIVA,][]{Chung2009} are found with H\,\textsc{i} discs truncated to within the stellar disc with the smallest $d_{\rm{HI}}/d_{\rm{opt}}\sim0.2$ \citep{Chung2009,Yoon2017}. In Hydra~I, we only find four H\,\textsc{i} detected cluster galaxies ($\sim10\%$ of the cluster population) with $d_{\rm{HI}}/d_{\rm{opt}}<1$ ($d_{\rm{HI}}/d_{\rm{opt}}=0.8$--0.9 of which two are only marginally resolved, i.e.\ $d_{\rm{HI}}<60$\,arcsec). We note that the two marginally resolved galaxies may not have their H\,\textsc{i} truncated within the optical disc (i.e.\ $d_{\rm{HI}}/d_{\rm{opt}}\geq1.0$) due to the synthesised beam induced bias in the inclination correction of the H\,\textsc{i} surface density (see Section~\ref{ss-sec:hi_diameter}). Although we do not find a similarly large population of severely truncated H\,\textsc{i} discs in Hydra~I, we can still see H\,\textsc{i} disc truncation as galaxies move from the field, through the infall region and into the cluster from the $d_{\rm{HI}}/d_{\rm{opt}}$ distributions and medians for the three populations (panel a of Figure~\ref{fig:histograms_diameter}). We also find that high-mass ($M_*\geq10^9\,\rm{M}_{\odot}$) cluster galaxies have smaller offsets from the gas fraction main sequence ($\Delta \rm{GFMS}$) than those in the infall or field. This indicates H\,\textsc{i} gas is being removed from cluster galaxies even though these galaxies are not yet gas-poor. The galaxies we observe with WALLABY are in an earlier stage of losing their H\,\textsc{i} gas reservoirs than the Virgo galaxies observed in the VIVA survey, which are gas-poor unlike the population we detect in Hydra~I. The smaller $d_{\rm{HI}}/d_{\rm{opt}}$ of infall and cluster galaxies is likely the result of a combination of environmental processes stripping gas from the H\,\textsc{i} discs (e.g.\ ram pressure and tidal stripping) and inflows, which would replenish gas, being cut off (e.g.\ starvation) as galaxies enter and traverse the cluster. While our data are not particularly sensitive to strong H\,\textsc{i} disc truncation, we are able to probe the early stages of gas removal as galaxies fall into the cluster. WALLABY is simply not sensitive to galaxies with severely truncated H\,\textsc{i} discs due to its sensitivity limit at the distance of Hydra~I.

Using galaxies within $<2.5R_{200}$ of Hydra~I from the same WALLABY data, \cite{Wang2021} investigated ram pressure stripping of H\,\textsc{i} in Hydra~I and classified the majority ($\sim75\%$) of galaxies within $R_{200}$ as candidates for experiencing ram pressure. Our analysis is in agreement with the results of \cite{Wang2021} and we conclude that ram pressure is contributing to the systematically smaller $d_{\rm{HI}}/d_{\rm{opt}}$ of cluster galaxies compared to our field population. Our infall population also has systematically smaller $d_{\rm{HI}}/d_{\rm{opt}}$ than the field. \cite{Wang2021} find that $\lesssim10\%$ of galaxies beyond $R_{200}$ to be ram pressure stripped candidates. This does not exclude ram pressure as the cause for the smaller $d_{\rm{HI}}/d_{\rm{opt}}$ of the infall population as our analysis probes small variations in the disc outskirts which are not included in the \cite{Wang2021} criteria for ram pressure stripping. \cite{Wang2021} find gas for which the ram pressure strength is greater than the galaxy's gravitational potential can be stripped on short timescales (i.e.\ $\lesssim200$\,Myr). Hence, ram pressure may have already removed the gas that could be stripped, producing the smaller $d_{\rm{HI}}/d_{\rm{opt}}$ for infall galaxies relative to the field. 

\begin{figure}
	\includegraphics[width=\columnwidth]{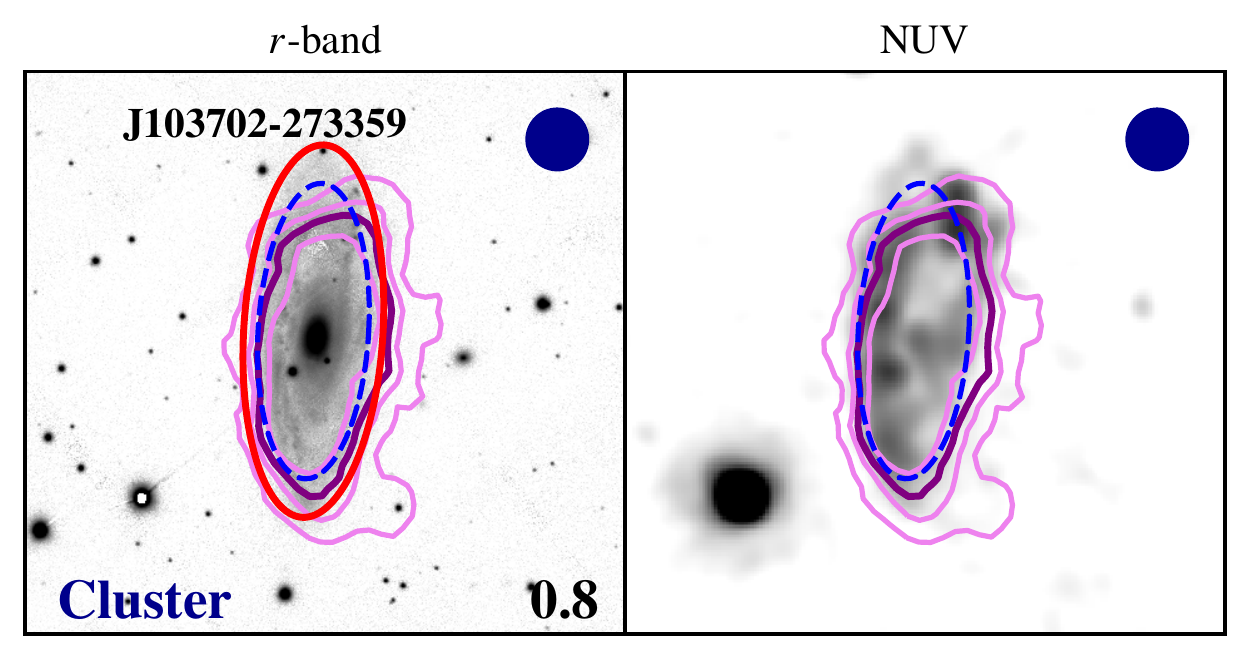}
    \caption{Similar to Figure~\ref{fig:galaxy_plate} showing the galaxy with the most truncated H\,\textsc{i} disc ($d_{\rm{HI}}/d_{\rm{opt}}\sim0.8$) in our sample: WALLABY J103702$-$273359 (NGC\,3312). The additional pink contours show H\,\textsc{i} column densities of 5, 20 and $50\times10^{19}\,\rm{cm}^{-2}$. \citet{Wang2021} find this galaxy to be experiencing ram pressure stripping, which is supported by the observed H\,\textsc{i} morphology and NUV map (i.e.\ NUV emission coincident with the compression of H\,\textsc{i} column density contours on the left (eastern) side, right panel).}
    \label{fig:single_plate}
\end{figure}

Figure~\ref{fig:single_plate} shows the PanSTARRS $r$-band and \textit{GALEX} NUV images with overlaid ellipses indicating the $r$-band and NUV disc diameters (solid red and dashed blue, respectively) and the purple contour showing the H\,\textsc{i} surface density of $1\,\rm{M}_{\odot}\,\rm{pc}^{-2}$ of the galaxy with the smallest $d_{\rm{HI}}/d_{\rm{opt}}=0.8$: WALLABY J103702$-$273359 (hereafter referred to by its NGC number: NGC\,3312). The projected distance of NGC\,3312 from the centre of Hydra~I is $r<0.1R_{200}$. For NGC\,3312, ram pressure stripping is likely the dominant mechanism responsible for the truncated H\,\textsc{i} disc due to the coincident NUV emission and compressed H\,\textsc{i} column density contours on the left (eastern) side of the galaxy. This interpretation is in agreement with the results of \cite{Wang2021}, who also find that NGC\,3312 to be experiencing ram pressure stripping.

\subsection{Star Formation}
\label{s-sec:sfms}

The H\,\textsc{i}-detected galaxies across each of the cluster, infall and field populations are predominantly star-forming. All three populations contain a small number of galaxies with SFRs more than $1\sigma=0.3$\,dex below or above the star forming main sequence, which shows the scatter of the star forming main sequence is found in all environments. Similar variations of potentially enhanced and suppressed star formation in individual galaxies have been found in the Virgo cluster \citep[e.g.][]{Koopmann2004a}. The majority of galaxies either currently undergoing quenching or already quenched in and around Hydra~I are not detected in H\,\textsc{i} with WALLABY. The ATLAS$^{\rm{3D}}$ H\,\textsc{i} survey detected four Virgo early-type galaxies in H\,\textsc{i} with H\,\textsc{i} masses of $M_{\rm{HI}}\sim10^7$--$10^{8.75}\,\rm{M}_{\odot}$ and placed upper limits of $M_{\rm{HI}}<10^8\,\rm{M}_{\odot}$ on H\,\textsc{i} non-detections \citep{Serra2012}. Assuming this is representative of the H\,\textsc{i} in cluster early-type galaxies, it is unsurprising that we do not detect these galaxies in Hydra~I, as these systems fall below the WALLABY detection limit at 61\,Mpc. The fraction of cluster galaxies that are experiencing/have experienced quenching is higher than that of the infall population as expected. Cluster galaxies will have experienced environmental processes capable of stripping or depleting their gas content over longer time scales, which in turn will lead to star formation quenching. 

A small number of cluster and infall H\,\textsc{i} non-detections and all the field non-detections lie on the star forming main sequence. Unlike the H\,\textsc{i} non-detections with large offsets from the main sequence, which likely have limited gas reservoirs, these galaxies will presumably still have significant quantities of H\,\textsc{i} and would be detected with deeper H\,\textsc{i} observations. \cite{Cortese2021} illustrate the decreasing sensitivity of blind H\,\textsc{i} surveys at increasing distance by determining which galaxies in the xGASS sample would be detected at different distances assuming the nominal WALLABY sensitivity ($\sigma=1.6$\,mJy\,beam$^{-1}$). At distances $\gtrsim60$\,Mpc, the majority of the passive population from xGASS falls below the WALLABY detection limit, in agreement with our results. We also conclude that H\,\textsc{i} stacking will be required to look for H\,\textsc{i} in quenching galaxies \citep[e.g.\ as done for the Coma Cluster,][]{Healy2021a}. The fact we that we do not find a reduction in the galaxy-wide star formation of the infall or cluster populations, but do find smaller H\,\textsc{i} discs, leads us to conclude that gas is mainly being stripped from the galaxy outskirts and is not yet affecting the inner star-forming disc. We also find that cluster galaxies have lower $\Delta \rm{GFMS}$ relative to the infall and field populations without a corresponding decrease in $\Delta \rm{SFMS}$ (Figures~\ref{fig:sfms_gfms} and \ref{fig:histograms_ms}). This indicates that there is a time delay between the removal of H\,\textsc{i} gas and star formation quenching. 

Our conclusions are supported by \cite{Li2020}, who find that cluster galaxies in the EAGLE simulation show H\,\textsc{i} gas loss while remaining on the star forming main sequence, as H\,\textsc{i} is preferentially lost from the galaxy outskirts. This suggests that there is a time delay between H\,\textsc{i} gas removal and quenching. \cite{Oman2016} and \cite{Oman2021} also show this using $N$-body simulations. Other cosmological simulations that properly account for environmental effects have been demonstrated to produce galaxies that fit this description \citep[e.g.][]{Stevens2017,Cora2018,Stevens2021} as well.

Focusing on the outskirts of our galaxy populations' discs (i.e.\ beyond the optical disc), we find that recent star formation as traced by the NUV emission is truncated in cluster galaxies compared with field galaxies (panel g of Figure~\ref{fig:histograms_diameter}). We also find that cluster galaxies have smaller H\,\textsc{i} to NUV disc size ratios, which explains the smaller discs of recent star formation, as these galaxies have smaller gas reservoirs in their outskirts. Truncation of the star-forming disc of cluster galaxies is also found in other clusters using different tracers (e.g.\ near-UV, H$\alpha$ and 24$\mu$m) of recent star formation \citep[e.g.][]{Cortese2012,Fossati2013,Finn2018}. The reduction in the size of the disc of recent star formation in cluster and infall galaxies supports our conclusion that H\,\textsc{i} is being stripped from the outskirts of these galaxies.

\section{Conclusions}
\label{sec:conclusion}

In this work, we take advantage of the wide-field capabilities of ASKAP using WALLABY pilot survey observations of the Hydra~I cluster covering a 60-square-degree field of view. We probe the effect of the environment by comparing uniformly measured H\,\textsc{i} properties in populations of cluster, infall and field galaxies out to $\sim5R_{200}$ from the cluster centre. We find that the H\,\textsc{i} detected fraction decreases from $\sim0.85$ (comparable to the field fraction of $\sim0.9$) to $\sim0.35$ at a projected distance of $r\sim1.5R_{200}$, indicating that the environment begins strongly affecting the H\,\textsc{i} content of galaxies at projected distances $r\lesssim1.5R_{200}$ from Hydra~I.

Previous studies that compared the H\,\textsc{i} of cluster and field galaxies over a similarly large field of view were limited to integrated properties from single dish surveys \citep[e.g.][]{Solanes2001}. Using the high spatial resolution of WALLABY (30\,arcsec), we measure the H\,\textsc{i} isodensity to  optical isophotal 23.5\,mag\,arcsec$^{-2}$ $r$-band diameter ratio, $d_{\rm{HI}}/d_{\rm{opt}}$, for the three galaxy populations and find that the median $d_{\rm{HI}}/d_{\rm{opt}}$ decreases from the field, through the infall regime, and further into the cluster. This indicates that environmental processes are stripping gas as galaxies enter the denser cluster environment and producing smaller H\,\textsc{i} discs. We do not find a population of severely truncated H\,\textsc{i} discs analogous to those found in the Virgo cluster \citep{Chung2009,Yoon2017}. At the distance of Hydra~I (61\,Mpc), WALLABY is not sensitive to galaxies with H\,\textsc{i} discs truncated to within the stellar disc with H\,\textsc{i} column densities $\lesssim10^{20}$\,cm$^{-2}$. Galaxies with severely truncated H\,\textsc{i} discs probably have H\,\textsc{i} masses below the sensitivity of WALLABY (e.g.\ $M_{\rm{HI}}<10^{8.4}\,\rm{M}_{\odot}$ for unresolved sources or $M_{\rm{HI}}<10^{8.7}\,\rm{M}_{\odot}$ for a typical resolved source with a stellar mass of $M_*\sim10^9\,\rm{M}_{\odot}$). Although our data are not sensitive to strong H\,\textsc{i} disc truncation, we can probe the first stages of gas removal and disc truncation.

We also investigate star formation in the three galaxy populations. We find that the distributions of H\,\textsc{i}-detected cluster, infall and field galaxies are all centred on the star forming main sequence. Cluster and infall galaxies that are undergoing quenching or are already quenched, which lie below the main sequence, are not detected in H\,\textsc{i} with WALLABY at a distance of 61\,Mpc. Using the NUV emission to trace recent star formation, we find that there is less recent star formation occurring beyond the optical disc of galaxies in the cluster than in the field as a result of smaller gas reservoirs in the outer discs of cluster galaxies. We conclude that H\,\textsc{i} gas is primarily being stripped from the outskirts of infalling galaxies and not yet affecting the gas reservoir of their inner star-forming discs.

These results demonstrate the transformational power of WALLABY to provide statistically significant galaxy samples to probe resolved galaxy properties in a uniform manner over a wide range of environments. The full WALLABY survey will enable similar analyses to that presented here of environment densities ranging from low-density, isolated galaxies to intermediate groups and high-density clusters using $\sim500\,000$ galaxies, of which $\sim5\,000$ will be spatially resolved \citep{Koribalski2020}. The WALLABY reference simulation \citep{Koribalski2020} predicts that, with (without) accounting for stripping, $\sim500$ (1\,500) spatially resolved galaxies detected with WALLABY reside in clusters $>6\times10^{14}\,\rm{M}_{\odot}$ (e.g.\ Virgo like) and $\sim1\,500$ (4\,500) reside in clusters $>3\times10^{14}\,\rm{M}_{\odot}$ (e.g.\ Hydra~I like). This will enable studies of resolved H\,\textsc{i} properties that were previously limited to integrated parameters from single-dish all-sky surveys (e.g.\ HIPASS). Further insights into the gas properties of galaxies not directly detected will be made possible through stacking of H\,\textsc{i} non-detections (e.g.\ to probe H\,\textsc{i} in gas-poor galaxies and galaxies below the star forming main sequence).

\section*{Acknowledgements}

We thank F.~Lelli, H.~Courtois, D.~Pomarede, A.~Boselli and F.~Bigiel for helpful discussions and comments. We also thank the anonymous referee for their comments.

This research was conducted by the Australian Research Council Centre of Excellence for All Sky Astrophysics in 3 Dimensions (ASTRO 3D), through project number CE170100013. 

The Australian SKA Pathfinder is part of the Australia Telescope National Facility which is managed by the Commonwealth Scientific and Industrial Research Organisation (CSIRO). Operation of ASKAP is funded by the Australian Government with support from the National Collaborative Research Infrastructure Strategy. ASKAP uses the resources of the Pawsey Supercomputing Centre. Establishment of ASKAP, the Murchison Radio-astronomy Observatory (MRO) and the Pawsey Supercomputing Centre are initiatives of the Australian Government, with support from the Government of Western Australia and the Science and Industry Endowment Fund. We acknowledge the Wajarri Yamatji as the traditional owners of the Observatory site. We also thank the MRO site staff. This paper includes archived data obtained through the CSIRO ASKAP Science Data Archive, CASDA (\url{http://data.csiro.au}). 

LC acknowledges support from the Australian Research Council Discovery Project and Future Fellowship funding schemes (DP210100337,FT180100066).

DO is a recipient of an Australian Research Council Future Fellowship (FT190100083) funded by the Australian Government.

JR acknowledge support from the State Research Agency (AEI-MCINN) of the Spanish Ministry of Science and Innovation under the grant `The structure and evolution of galaxies and their central regions' with reference PID2019-105602GB-I00/10.13039/501100011033.

LVM acknowledges financial support from the State Agency for Research of the Spanish Ministry of Science, Innovation and Universities through the `Center of Excellence Severo Ochoa' awarded to the Instituto de Astrof\'isica de Andaluc\'ia (SEV-2017-0709), from the grant RTI2018-096228-B-C31 (Ministry of Science, Innovation and Universities/State Agency for Research/European Regional Development Funds, European Union), from the grant RED2018-102587-T (Spanish Ministry of Science, Innovation and Universities/State Agency for Research) and IAA4SKA P18-RT-3082 (Regional Government of Andalusia).

AB acknowledges support from the Centre National d’Etudes Spatiales (CNES), France.

ARHS acknowledges receipt of the Jim Buckee Fellowship at ICRAR-UWA.

SHO acknowledges support from the National Research Foundation of Korea (NRF) grant funded by the Korea government (Ministry of Science and ICT: MSIT) (No. NRF-2020R1A2C1008706).

This project has received funding from the European Research Council (ERC) under the European Union’s Horizon 2020 research and innovation programme (grant agreement no. 679627; project name FORNAX). 

This research has made use of the NASA/IPAC Extragalactic Database (NED) which is operated by the Jet Propulsion Laboratory, California Institute of Technology, under contract with the National Aeronautics and Space Administration.

The Pan-STARRS1 Surveys (PS1) and the PS1 public science archive have been made possible through contributions by the Institute for Astronomy, the University of Hawaii, the Pan-STARRS Project Office, the Max-Planck Society and its participating institutes, the Max Planck Institute for Astronomy, Heidelberg and the Max Planck Institute for Extraterrestrial Physics, Garching, The Johns Hopkins University, Durham University, the University of Edinburgh, the Queen's University Belfast, the Harvard-Smithsonian Center for Astrophysics, the Las Cumbres Observatory Global Telescope Network Incorporated, the National Central University of Taiwan, the Space Telescope Science Institute, the National Aeronautics and Space Administration under Grant No. NNX08AR22G issued through the Planetary Science Division of the NASA Science Mission Directorate, the National Science Foundation Grant No. AST-1238877, the University of Maryland, Eotvos Lorand University (ELTE), the Los Alamos National Laboratory, and the Gordon and Betty Moore Foundation.

This publication makes use of data products from the Wide-field Infrared Survey Explorer (WISE), which is a joint project of the University of California, Los Angeles, and the Jet Propulsion Laboratory/California Institute of Technology, funded by the National Aeronautics and Space Administration.

This work is based in part on observations made with the Galaxy Evolution Explorer (\textit{GALEX}). \textit{GALEX} is a NASA Small Explorer, whose mission was developed in cooperation with the Centre National d'Etudes Spatiales (CNES) of France and the Korean Ministry of Science and Technology. \textit{GALEX} is operated for NASA by the California Institute of Technology under NASA contract NAS5-98034.

\section*{Data Availability}

The derived galaxy properties used in this work are available as supplementary material. We show the first five rows of the tables for the H\,\textsc{i} detections and non-detections in Appendix~\ref{appendix:data_tables}. The full 36-beam, 30-square-degree H\,\textsc{i} spectral line cubes (SBIDs 10269, 10609, 10612 and 10626), which are combined to create the full sensitivity 60-square-degree field, are available from the CSIRO ASKAP Science Data Archive \citep[CASDA,][]{Chapman2015,Huynh2020} using the DOI \url{https://doi.org/10.25919/5f7bde37c20b5}. PanSTARRS \citep{Chambers2016,Flewelling2016} $g$- and $r$-band imaging is available through the PanSTARRS cutout server \url{https://ps1images.stsci.edu/cgi-bin/ps1cutouts}. \textit{GALEX} DR6$+$7 \citep{Bianchi2017} NUV band imaging is available from \url{http://galex.stsci.edu/data}. unWISE band imaging data is available for W1 from \url{http://unwise.me/data/neo6/unwise-coadds/fulldepth} and for W3 and W4 from \url{http://unwise.me/data/allwise/unwise-coadds/fulldepth} \citep{Lang2014,Meisner2017}.



\bibliographystyle{mnras}
\bibliography{master} 



\appendix

\section{Data Tables}
\label{appendix:data_tables}

We show the first five rows of the online supplementary data tables containing the parameters used in this work for the galaxies detected and not detected in H\,\textsc{i} in Tables~\ref{table:param_hi_detections} and \ref{table:param_hi_nondetections}, respectively.

\begin{table*}
	\centering
	\caption{Measured and derived properties for galaxies detected in H\,\textsc{i} with WALLABY. The full table is available as supplementary material.}
	\label{table:param_hi_detections}
	\begin{tabular}{lcccccccccr}
	    \hline
		WALLABY ID  & Environment & $D_{\rm{L}}$ & $\frac{r}{R_{200}}$ & $\frac{\Delta v}{\sigma_{\rm{disp}}}$ & $\log\left(\frac{M_*}{\rm{M}_{\odot}}\right)$ & $\log\left(\frac{M_{\rm{HI}}}{\rm{M}_{\odot}}\right)$   & $d_{\rm{HI}}$ & $d_{\rm{opt}}$ & $d_{\rm{NUV}}$  & $\log\left(\frac{\rm{SFR}}{\rm{M}_{\odot}\,\rm{yr}^{-1}}\right)$  \\
		 &  & [Mpc] &  &  &  &  & [arcsec] & [arcsec] & [arcsec] &   \\
		\hline
		J100336-262923 & field & 13.3 & 5.3 & 4.3 & 6.4  & 8.0 & 98  & 10  & 8   & ---      \\
		J100342-270137 & field & 14.2 & 5.2 & 4.2 & 8.5  & 9.4 & 425 & 88  & 164 & $-0.87$  \\
		J100351-263707 & field & 12.9 & 5.3 & 4.3 & 7.2  & 8.6 & 151 & 28  & 44  & $-1.69$  \\
		J100351-273417 & field & 41.5 & 5.2 & 1.5 & 10.4 & 9.9 & 148 & 126 & 111 & $0.25$   \\
		J100426-282638 & field & 16.0 & 5.1 & 4.0 & 9.0  & 9.4 & 332 & 150 & 162 & $-0.75$  \\
		\hline
	\end{tabular}
\end{table*}

\begin{table*}
	\centering
	\caption{Measured and derived properties for galaxies in the 6dFGS catalogue that are not detected in H\,\textsc{i} with WALLABY. The full table is available as supplementary material.}
	\label{table:param_hi_nondetections}
	\begin{tabular}{lcccccccccr}
	    \hline
		6dFGS ID  & Environment & $D_{\rm{L}}$ & $\frac{r}{R_{200}}$ & $\frac{\Delta v}{\sigma_{\rm{disp}}}$ & $\log\left(\frac{M_*}{\rm{M}_{\odot}}\right)$ & $\log\left(\frac{M_{\rm{HI}}}{\rm{M}_{\odot}}\right)$   & $d_{\rm{HI}}$ & $d_{\rm{opt}}$ & $d_{\rm{NUV}}$  & $\log\left(\frac{\rm{SFR}}{\rm{M}_{\odot}\,\rm{yr}^{-1}}\right)$  \\
		 &  & [Mpc] &  &  &  &  & [arcsec] & [arcsec] & [arcsec] &   \\
		\hline
		g1006222-264958 & field  & 68.3  & 4.8 & 1.2 & 8.9  & $<8.7$ & $<13$ & 19  & 14   & $-0.38$  \\
		g1010470-285406 & field  & 63.1  & 4.2 & 0.7 & 10.2 & $<8.9$ & $<16$ & 51  & ---  & ---      \\
		g1019338-264156 & field  & 36.0  & 2.8 & 2.0 & 10.6 & $<8.4$ & $<9$  & 150 & 72   & ---      \\
		g1023024-260448 & infall & 54.1  & 2.4 & 0.2 & 9.4  & $<8.6$ & $<11$ & 23  & 30   & $-0.20$  \\
		g1025185-295118 & infall & 59.6  & 2.4 & 0.3 & 8.8  & $<8.6$ & $<11$ & 22  & 18   & $-1.01$  \\
		\hline
	\end{tabular}
\end{table*}


\bsp	
\label{lastpage}
\end{document}